\documentclass[%
superscriptaddress,
 amsmath,amssymb,
 aps,
 twocolumn,
prm,
]{revtex4-2}

\usepackage{graphicx}
\usepackage{dcolumn}
\usepackage{bm}
\usepackage{multirow}
\usepackage{hyperref}

\makeatletter
\newsavebox{\@brx}
\newcommand{\llangle}[1][]{\savebox{\@brx}{\(\m@th{#1\langle}\)}%
  \mathopen{\copy\@brx\mkern2mu\kern-0.9\wd\@brx\usebox{\@brx}}}
\newcommand{\rrangle}[1][]{\savebox{\@brx}{\(\m@th{#1\rangle}\)}%
  \mathclose{\copy\@brx\mkern2mu\kern-0.9\wd\@brx\usebox{\@brx}}}
\makeatother

\begin{document}

\title{Superconductor discovery in the emerging paradigm of Materials Informatics}

\author{Huan Tran}
\email{huan.tran@mse.gatech.edu}
\affiliation{School of Materials Science \& Engineering, Georgia Institute of Technology, 771 Ferst Dr. NW, Atlanta, GA 30332, USA}
\author{Hieu-Chi Dam}
\affiliation{Japan Advanced Institute of Science and Technology, 1-1 Asahidai, Nomi, Ishikawa 923-1292, Japan}
\author{Christopher Kuenneth}
\affiliation{Faculty of Engineering Science, University of Bayreuth, Universitätsstr. 30, 95447 Bayreuth, Germany}
\author{Tuoc N. Vu}
\affiliation{Institute of Engineering Physics, Hanoi University of Science \& Technology, 1 Dai Co Viet Rd., Hanoi 10000, Vietnam}
\author{Hiori Kino}
\affiliation{Research Center for Electronic and Optical Materials, National Institute for Materials Science, 1-1 Namiki, Tsukuba, Ibaraki 305-0044, Japan}

\date{\today}

\begin{abstract}
The last two decades have witnessed a tremendous number of computational predictions of hydride-based (phonon-mediated) superconductors, mostly at extremely high pressures, i.e., hundreds of GPa. These discoveries were heavily driven by Migdal-\'{E}liashberg theory (and its first-principles computational implementations) for electron-phonon interactions, the key concept of phonon-mediated superconductivity. Dozens of predictions were experimentally synthesized and characterized, triggering not only enormous excitement in the community but also some debates. In this Article, we review the computational-driven discoveries and the recent developments in the field from various essential aspects, including the theoretical, computational, and, specifically, artificial intelligence (AI)/machine learning (ML) based approaches emerging within the paradigm of materials informatics. While challenges and critical gaps can be found in all of these approaches, AI/ML efforts specifically remain in its infant stage for good reasons. However, opportunities exist when these approaches can be further developed and integrated in concerted efforts, in which AI/ML approaches could play more important roles.
\end{abstract}

\pacs{}
\maketitle



\section{Introduction}
\begin{figure}[t]
\centering
\vspace{2mm}
\includegraphics[width=0.95\linewidth]{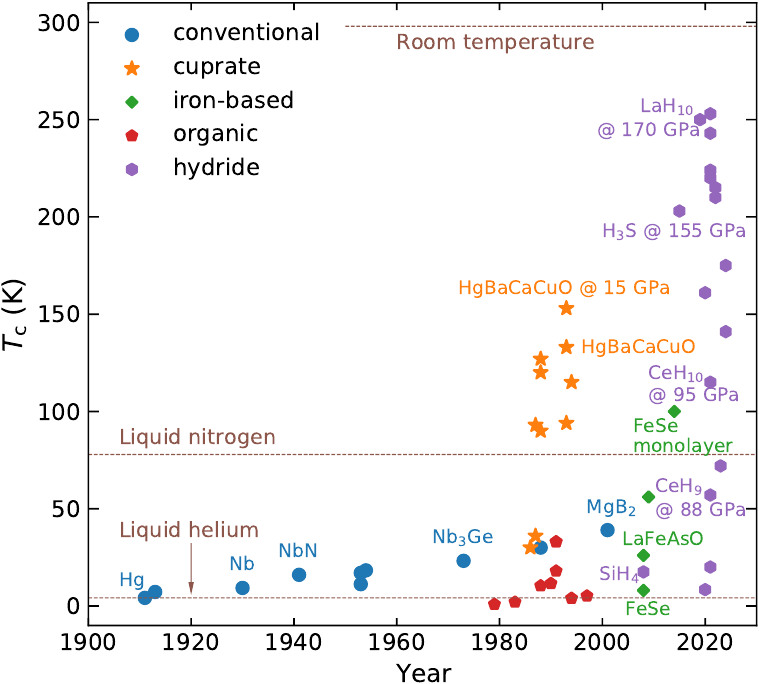}
\caption{The time evolution of $T_{\rm c}$ for some classes of superconductors.}\label{fig:Tchist}
\end{figure}

Superconductivity was discovered by H. Kamerlingh Onnes in 1911 \cite{onnes1911}. When cooling some ordinary substances such as mercury and lead down to a critical temperature $T_{\rm c}$ near absolute zero, Onnes found that their electrical resistance $R$ completely disappears, signaling that they can carry a current indefinitely without losing any energy. In this state, a superconducting material is a perfect diamagnet, i.e., it completely repels external magnetic fields in the phenomenon known as the Meissner effect \cite{meissner1933neuer}. Although $T_{\rm c}$ discovered for the first superconductors is generally very low, e.g., just a few K \cite{smith1935superconductivity, eisenstein1954superconducting,matthias1963superconductivity}, the discovery of superconductivity triggered not only the development of a new branch of condensed matter physics \cite{london1948problem, landau1950theory, frohlich1954electrons, frohlich1954theory, bardeen1955electron, bardeen1955theory, cooper1956bound, bardeen1957microscopic, bardeen1973electron, schrieffer2018theory, tinkham2004introduction, ginzburg2004superconductivity, carbotte1990properties, Eliashberg, Migdal1958, chubukov2020eliashberg, marsiglio2020eliashberg,allen1983theory, pellegrini2024ab, kresin1993mechanisms, stewart2017unconventional, anderson1997theory, blatter1994vortices, lee2006doping, orenstein2000advances} but also significant interest from society \cite{johnson2023superconductor,cartlidge2015superconductivity,jin2023hopes} in the century to come. The ultimate research goal is a material that ``superconducts'' electricity under ambient conditions, i.e., at the room temperature and above ($T_{\rm c}\geq 300$ K) and at atmospheric pressure ($P\simeq 0.1$ GPa). If such a material could be discovered, it may unlock many technologies previously confined to science fiction, potentially transforming human civilization as we know it \cite{little1965superconductivity, sleight1995room}. Before the 21$^{\rm st}$ century, numerous new superconductors were discovered \cite{matthias1952search,muller1987discovery,bednorz1986possible,bednorz1988perovskite, chu1987evidence, cava1987bulk, wu1987superconductivity, cava1987bulk2, hazen1988superconductivity, schilling1993superconductivity, chu1993superconductivity, nagamatsu2001superconductivity,kamihara2006iron, kamihara2008iron, hosono2015exploration, hosono2015iron}, pushing the record of $T_{\rm c}$ to $\simeq 150$ K \cite{schilling1993superconductivity, chu1993superconductivity} under (near-)ambient pressure (see Fig. \ref{fig:Tchist}). This period was largely propelled by experimental efforts, often performed in combinations with profound physics expertise and intuition.

Concurrently with experimental advancements, a new branch of theoretical condensed matter physics emerged, dedicating to the understanding of superconductivity, ``a manifestation of the quantum world on a macroscopic scale'' \cite{marsiglio2020eliashberg}, from the atomic level \cite{london1948problem, landau1950theory, frohlich1954electrons, frohlich1954theory, bardeen1955electron, bardeen1955theory, cooper1956bound,bardeen1957microscopic,bardeen1973electron,schrieffer2018theory, tinkham2004introduction, ginzburg2004superconductivity,carbotte1990properties, Eliashberg,Migdal1958, chubukov2020eliashberg, marsiglio2020eliashberg,allen1983theory,kresin1993mechanisms, stewart2017unconventional, anderson1997theory}. The first solid microscopic theory of superconductivity was developed by Bardeen, Cooper, and Schrieffer (BCS) in 1957 \cite{bardeen1957microscopic, bardeen1973electron, cooper1956bound}. Long story short, lattice phonons can mediate a net attractive interaction between certain pairs of electrons in the neighborhood of the Fermi surface, fusing them into a bound state called Cooper pair \cite{cooper1956bound}. Below $T_{\rm c}$, the Cooper pairs can form a robust quantum condensation that can flow without dissipations. Some years later, Migdal-\'{E}liashberg theory was developed \cite{Eliashberg,Migdal1958, chubukov2020eliashberg, marsiglio2020eliashberg,allen1983theory,pellegrini2024ab}, providing a more complete, truly many-body approach for the simplified model of instantaneous electron-phonon (EP) interactions in the BCS theory. From another end, theoretical endeavors \cite{kresin1993mechanisms, stewart2017unconventional, anderson1997theory, blatter1994vortices, lee2006doping, orenstein2000advances} have expanded far beyond the phonon-mediated pairing mechanism, which applies to about one third of known (conventional) superconductors \cite{hirsch2022superconducting}. For the remaining two-thirds of (unconventional) superconductors, where this pairing mechanism does not apply, dozens of new mechanisms were suggested, involving possible roles of screened Coulomb interactions \cite{kohn1965new}, polarons \cite{mott1993polaron}, anyons \cite{chen1989anyon}, Majorana fermions \cite{beenakker2013search}, topology \cite{leijnse2012introduction}, spin fluctuations \cite{scalapino1999superconductivity}, resonating valence-bond states \cite{anderson1987resonating}, and more. Unconventional superconductors are arranged into numerous classes, e.g., cuprates \cite{bednorz1986possible}, iron-based \cite{kamihara2006iron, kamihara2008iron, hosono2015exploration, hosono2015iron}, heavy fermion materials \cite{petrovic2001heavy, steglich1984heavy}, and organic materials \cite{jerome1991physics, saito2011organic, ardavan2011recent, lebed2008physics}, each of them may be associated with one or more pairing mechanisms. Theoretically, the proposed mechanisms impose no inherent limits on $T_{\rm c}$ -- in fact, the Kohn-Luttinger theorem \cite{kohn1965new} allows for arbitrarily small $T_{\rm c}$. After all, superconductivity remains one of the most challenging and enigmatic topics of physics, with much yet to be understood.

Starting from the 2000s, first-principles computational approaches and the required infrastructure have progressed \cite{OganovBook, oganov2019structure, NeedsReview:CMD, Zhang:NRM17, xu2022materials, DFPT,giustino2017electron} to the point where they can provide, within the framework of the Migdal-\'{E}liashberg theory, some valuable guidance for the searches of new phonon-mediated superconductors \cite{pickett2022room, zurek2019high, sun2023clathrate, hilleke2022tuning, gao2021superconducting, zhao2023superconducting, boeri2019road, pickard2020superconducting, wang2012superconductive, duan2014pressure, errea2020quantum, Alexey:CaB, Huan:Mg-Si, li2015pressure, liu2017potential, peng2017hydrogen, kvashnin2018high, zhang2022design, boeri20212021, yazdani2022artificial, ribeiro2004possible, saniz2004electronic, di2021bh, zhang2024high, sanna2024prediction}. Two classes of first-principles-based methods that are critical for superconductor discovery are structure prediction methods \cite{OganovBook, oganov2019structure} and density functional perturbation theory (DFPT) \cite{DFPT,giustino2017electron}. The former is used to explore the configuration space, while the latter is used to approximate the EP interactions. Equipped with these tools, ones may start from a hypothetical chemical formula, predicting the lowest-enthalpy atomic structures at specific pressures, examining their thermodynamic and dynamical stability, evaluating the EP interactions, and, finally, estimating their $T_{\rm c}$. This generic workflow (shown in Fig. \ref{fig:flow} (a)) was used predominantly during the last two decades, leading to thousands of predictions for high-$T_{\rm c}$ materials \cite{pickett2022room, wang2012superconductive, duan2014pressure, zurek2019high,sun2023clathrate, hilleke2022tuning,errea2020quantum, gao2021superconducting, zhao2023superconducting, boeri20212021, yazdani2022artificial, boeri2019road, pickard2020superconducting, Alexey:CaB, Huan:Mg-Si, li2015pressure, liu2017potential, peng2017hydrogen, kvashnin2018high, zhang2022design}. Dozens of them, including SiH$_4$ \cite{eremets2008superconductivity}, H$_3$S \cite{Drozdov15}, LaH$_{10}$ \cite{somayazulu2019evidence,drozdov2019superconductivity}, ThH$_{10}$ \cite{semenok2020superconductivity}, BaH$_{12}$ \cite{chen2021synthesis}, YH$_6$ \cite{troyan2021anomalous, kong2021superconductivity}, YH$_9$ \cite{kong2021superconductivity}, CeH$_9$ and CeH$_{10}$ \cite{chen2021high}, CaH$_6$ \cite{ma2022high,li2022superconductivity}, and LaBeH$_8$ \cite{song2023stoichiometric} were synthesized and tested experimentally (see Table \ref{table:disc} for a summary), elevating $T_{\rm c}$ to a record of $\simeq 250$ K, but at the cost of ultra-high pressures of $P \simeq 100 - 200$ GPa. Despite these achievements, critical gaps remain in various stages of the workflow, possibly impacting its predictive power, which is one of the points raised \cite{hirsch2022superconducting} in the on-going debates in the field \cite{hirsch2021unusual, wang2021absence, hirsch2021nonstandard, gubler2022missing, eremets2022high, hirsch2022superconducting, xing2023observation, chen2023muted}.

Fig. \ref{fig:Tchist} provides a snapshot of the $T_{\rm c}$ evolution during the last century. New discoveries, labeled as ``hydride superconductors'' in Fig. \ref{fig:Tchist} and summarized in Table \ref{table:disc}, were synthesized and tested experimentally while some of them were reproduced independently. The highest-$T_{\rm c}$ conventional superconductor is MgB$_2$ with $T_{\rm c}\simeq 39$ K \cite{nagamatsu2001superconductivity}, while unconventional superconductors raise $T_{\rm c}$ to $\simeq 150$ K \cite{schilling1993superconductivity, chu1993superconductivity}. The discoveries of hydride superconductors push the boundary to $\simeq 250$ K at hundreds of GPa. Some points should be noted on them. First, the discoveries are heavily driven by the Migdal-\'{E}liashberg theory-based first-principles computational workflow, suggesting the raising role of this non-experimental method. Second, hydride superconductors are predominant \cite{pickard2020superconducting}, perhaps because the field is strongly inspired by Ashcroft, who, in 2004, predicted \cite{ashcroft2004hydrogen, ashcroft2004bridgman} that high-$T_{\rm c}$ superconductivity may be found in hydrogen-dominant alloys, probably at high pressure $P$. The main rationale is that such materials may feature high vibrational frequencies involving the hydrogen atoms, thus enhancing the electron-phonon interactions. Finally, some extraordinary experimental claims were retracted in the last several years after facing valid concerns from the community \cite{hirsch2021unusual, wang2021absence, hirsch2021nonstandard, gubler2022missing, eremets2022high, hirsch2022superconducting, xing2023observation, chen2023muted}. We believe that transparency and healthy debates are important for the scientific integrity of all fields, including superconductivity.

The upsurge of computation-inspired superconductor discoveries \cite{pickett2022room, zurek2019high, sun2023clathrate, hilleke2022tuning, gao2021superconducting, zhao2023superconducting, boeri20212021, boeri2019road, pickard2020superconducting, wang2012superconductive, duan2014pressure, errea2020quantum, yazdani2022artificial, Alexey:CaB, Huan:Mg-Si, li2015pressure, liu2017potential, peng2017hydrogen, kvashnin2018high, zhang2022design} occurs almost simultaneously with the emergence of materials informatics \cite{Rampi:ML, agrawal2016perspective, jain2016new, draxl2018nomad, draxl2019nomad, draxl2020big, Luca:descriptor, damewood2023representations}, a new frontier of materials research. Within this paradigm, artificial intelligence (AI)/machine-learning (ML) based methods are developed to learn past data, ultimately accelerating the understanding, discovering, and designing of new materials. Given their nature, AI/ML approaches may, in principle, complement any physics-based experimental and computational methods that can produce reliable data. Despite considerable efforts \cite{yazdani2022artificial, boeri20212021, gu2019superconducting, yang2021hard, dolui2024feasible,ishikawa2024evolutionary, ferreira2023search, lucrezi2023quantum, xie2019functional, xie2022machine, isayev2015materials, hamidieh2018data, stanev2018machine, zeng2019atom, konno2021deep, choudhary2022designing, cerqueira2024sampling, tran2023machine, zhang2022machine, matsumoto2019acceleration, pereti2023individual, ishikawa2019materials, hutcheon2020predicting, shipley2021high, wines2023high, hai2023superconductivity, song2021high, le2020critical, garcia2021prediction, stanev2021artificial, raviprasad2022tree, revathy2022random, bai2024unveiling, roter2020predicting, roter2022clustering, yamaguchi2020sc, yamaguchi2022superconductivity, foppiano2019proposal, foppiano2021supermat, foppiano2023automatic, sommer20233dsc}, the development of materials informatics approaches for superconductor discovery remains in an early stage \cite{tran2023machine, bai2024unveiling}. Looking at other branches of materials informatics where concerted efforts with traditional physics-based methods are blossoming, it is obvious that superconductor discovery may be greatly benefited when the informatics-related developments could gain more momentum.

In this Article, we will review the recent computation-driven discoveries of the hydride superconductors with an eye to the emerging paradigm of materials informatics, which can, in principle, be useful for the future discovery of superconductors in any classes. For this goal, Sec. \ref{sec:methods} is devoted to the essential background of the hydride supderconductor discovery, including the Migdal-\'{E}liashberg theory, the first-principles computations of phonon-mediated superconductivity-related properties, and the computational workflow used widely for the discoveries. Sec. \ref{sec:methods} also includes a minimal coverage of some experimental methods needed for the discussions, although the main theme of this Article is non-experimental. Then, in Sec. \ref{sec:disc}, major computation-driven experimental discoveries that were reproduced and confirmed at some levels are reviewed. Next, a simplified introduction of materials informatics is given in Sec. \ref{sec:ML}, followed by a review on the materials informatics works that have been completed in this field. Finally, we offer in Sec. \ref{sec:challenge} some opinions on the remaining challenges and some forward-looking next steps, specifically in terms of what materials informatics can do to further contribute to the future of superconductor discovery.

\section{Fundamentals of superconductor discovery}\label{sec:methods}
\subsection{Migdal-\'{E}liashberg theory}\label{sec:theory}

In the BCS theory \cite{frohlich1954electrons, bardeen1957microscopic}, the phonon-mediated interaction between certain pairs of electrons residing within an energy cutoff $\omega_{\rm c}$ from the Fermi surface is assumed to be instantaneous and constant, while vanishing beyond $\omega_{\rm c}$. This simplified picture does not include enough the physics of interacting electron-phonon systems. Migdal-\'{E}liashberg theory is more completed \cite{Eliashberg,Migdal1958, allen1983theory, marsiglio2020eliashberg, chubukov2020eliashberg,pellegrini2024ab}, taking into account the retarded nature of the EP interactions, considering the first order of vertex corrections \cite{chubukov2020eliashberg}, while relying on Migdal theorem \cite{Migdal1958} to treat the damping of the excitations. Central to this theory are the anisotropic non-linear integral \'{E}liashberg equations involving the Matsubara gap $\Delta ({\bf k}, i\omega_n)$ and the renormalization factor $Z({\bf k}, i\omega_n)$. On the imaginary axis, they are \cite{margine2013anisotropic}
\begin{eqnarray}\label{Eq:El1}
Z({\bf k}, i\omega_n) =  1+\frac{\pi T}{N_{\rm F}\omega_n}\sum_{{\bf k}',n'}\frac{\omega_{n'}}{\sqrt{\omega^2_{n'}+\Delta^2({\bf k},i\omega_{n'})}} \nonumber \\
\times \lambda({\bf k}, {\bf k}', n-n'),
\end{eqnarray}
and
\begin{eqnarray}\label{Eq:El2}
Z({\bf k}, i\omega_n)\Delta({\bf k}, i\omega_n) = \frac{\pi T}{N_{\rm F}} \sum_{{\bf k}',n'}\frac{\Delta(i\omega_{n'})}{\sqrt{\omega^2_{n'}+\Delta^2({\bf k},i\omega_{n'})}} \nonumber \\
\times \left[\lambda({\bf k}, {\bf k}', n-n') - N_{\rm F}V({\bf k}-{\bf k}') \right].
\end{eqnarray}
Here, $N_{\rm F}$ is the density of state (DOS) at the Fermi level $\epsilon_{\rm F}$, $i\omega_n = i\pi T(2n-1)$ is the $n^{\rm th}$ Matsubara frequency where $n=0, \pm 1, \pm 2, ...$, $T$ is the temperature, and $V({\bf k}-{\bf k}')$ is the screened Coulomb interaction between the electronic states ${\bf k}$ and ${\bf k}'$. Furthermore,
\begin{eqnarray}\label{eq:lambda0}
\lambda({\bf k}, {\bf k}', n-n') = 2\int_0^\infty d\omega\frac{\omega}{\omega^2+(\omega_n-\omega_{n'})^2}\nonumber \\
\times \alpha^2F({\bf k}, {\bf k}', \omega)
\end{eqnarray}
is the EP coupling, determined from the EP spectral function
\begin{equation}\label{alpha2f_aniso}
\alpha^2F({\bf k}, {\bf k}', \omega) = N_{\rm F}\sum_{\nu} \vert g^{\nu}_{\bf k, k'}\vert^2 \delta(\omega-\omega^\nu_{\bf k-k'}).
\end{equation}
In Eq. \ref{alpha2f_aniso}, $g^{\nu}_{\bf k, k'}$ is the EP matrix elements, $\nu$ is the polarization index of the phonon with frequency $\omega$, and $\delta$ is the delta-Dirac function. When the anisotropy of the Fermi surface is weak, the isotropic spectral function
\begin{equation}\label{eq:a2f}
\alpha^2F(\omega) = \frac{1}{N_{\rm F}^2}\sum_{\bf k,k'}\alpha^2F({\bf k}, {\bf k}', \omega) \delta(\varepsilon_{\bf k}) \delta(\varepsilon_{\bf k'})
\end{equation}
can be used to suppress the $\bf k$ dependence in Eqs. (\ref{Eq:El1}), (\ref{Eq:El2}), and (\ref{eq:lambda0}), yielding two isotropic \'{E}liashberg equations and an isotropic EP coupling $\lambda(n-n')$. Within this approximation, the BCS gap equation can be reproduced by setting $\lambda(n-n') = \lambda$ for both $|\omega_{n}|, |\omega_{n'}| < \omega_{\rm c}$, and $0$ otherwise \cite{carbotte1990properties}, where
\begin{equation}\label{eq:lambda}
\lambda = 2\int_0^\infty d\omega\frac{\alpha^2F(\omega)}{\omega}
\end{equation}
is the dimensionless isotropic EP coupling $\lambda(n-n')$ at $n=n'$ that will be used extensively in the literature.

One way to examine the superconductivity of a material is to solve the \'{E}liashberg equations (\ref{Eq:El1}) and (\ref{Eq:El2}) self-consistently \cite{margine2013anisotropic, ponce2016epw,giustino2007electron} at multiple values of $T$ for a non-trivial solution $\Delta({\bf k}, i\omega_n)$, and the highest $T$ at which $\Delta({\bf k}, i\omega_n)\neq 0$ defines $T_{\rm c}$. In this procedure, two main inputs are the spectral function $\alpha^2F({\bf k}, {\bf k}', \omega)$ and a dimensionless Coulomb pseudopotential $\mu^*$, introduced as a treatment for the last term of Eq. (\ref{Eq:El2}). It is defined by \cite{margine2013anisotropic, ponce2016epw, morel1962calculation}
\begin{equation}\label{eq:mu1}
       \mu^* = \frac{N_{\rm F}\llangle V({\bf k}-{\bf k}')\rrangle}{1+N_{\rm F}\llangle V({\bf k}-{\bf k}')\rrangle \ln(\epsilon_{\rm F}/\omega_{\rm c})}
\end{equation}
where $\llangle ... \rrangle$ stands for the double average of the screened Coulomb interaction over ${\bf k}$ and ${\bf k}'$ on the Fermi surface. With $\mu^*$ defined in Eq. \ref{eq:mu1} and precomputed, the factor of $\left[\lambda({\bf k}, {\bf k}', n-n') - N_{\rm F}V({\bf k}-{\bf k}') \right]$ becomes $\left[\lambda({\bf k}, {\bf k}', n-n') - \mu^* \right]$ before Eqs. (\ref{Eq:El1}) and (\ref{Eq:El2}) are solved numerically \cite{margine2013anisotropic, ponce2016epw}.

To bypass the cumbersome step of solving the \'{E}liashberg equations (\ref{Eq:El1}) and (\ref{Eq:El2}), in 1968, McMillan \cite{McMillanTc} started from some solutions of these equations to develop a direct, empirical formula for $T_{\rm c}$, given that the isotropic spectral function $\alpha^2F(\omega)$ is known. It was then modified by Allen and Dynes \cite{dynes1972mcmillan, AllenTc} in 1975 to be
\begin{equation}\label{ADM0}
T_{\rm c} = \frac{\omega_{\log}}{1.2}\exp\left[-\frac{1.04(1+\lambda)}{\lambda-\mu^*(1+0.62\lambda)}\right]
\end{equation}
where
\begin{equation}\label{eq:omlog}
\omega_{\log} = \exp\left[\frac{2}{\lambda}\int_0^\infty d\omega\ln(\omega)\frac{\alpha^2F(\omega)}{\omega}\right]
\end{equation}
is the first logarithmic moment of $\alpha^2F(\omega)$. In the McMillan approach, $\mu^*$ is an empirical parameter typically ranging from 0.10 to 0.20 \cite{ashcroft2004hydrogen, ashcroft2004bridgman}. The empirical formula (\ref{ADM0}) is good for $\lambda\leq 1.5$ while additional empirical parameters are needed for larger values \cite{AllenTc}. Recently, there are some efforts \cite{xie2019functional, xie2022machine} to derive alternative formulae of (\ref{ADM0}), and they will be reviewed in Sec. \ref{sec:ml3}. Thanks to its simplicity, the McMillan formula (\ref{ADM0}) and some related versions \cite{AllenTc} were used extremely widely in the literature to estimate $T_{\rm c}$, given computed $\alpha^2F(\omega)$.

The domain of applicability is an essential aspect of a mathematical model. According to Migdal theorem \cite{Migdal1958}, the Migdal-\'{E}liashberg theory is believed \cite{Eliashberg} to be valid if the phonon energy scale is much smaller than the electronic energy scale, i.e., $\lambda\omega_{\rm c}/\epsilon_{\rm F} \ll 1$, even in the strong coupling regime, i.e., $\lambda \geq 1$. Subsequent examinations \cite{esterlis2018breakdown, alexandrov2001breakdown, hague2008breakdown, benedetti1998holstein, yuzbashyan2022breakdown, dynes1986breakdown} suggest a more complex picture. Among others, a common conclusion of these works is that the Migdal-\'{E}liashberg theory may be inaccurate when $\lambda$ exceeds a certain value, which can be 0.4 \cite{esterlis2018breakdown}, 1.0 \cite{alexandrov2001breakdown,  hague2008breakdown}, 1.3 \cite{benedetti1998holstein}, or 3.7 \cite{yuzbashyan2022breakdown}. Given that $\lambda \geq 1.0$ in most of the published reports employing this theory to predict $T_{\rm c}$ \cite{pickett2022room, zurek2019high, sun2023clathrate, hilleke2022tuning, gao2021superconducting, zhao2023superconducting, boeri2019road,pickard2020superconducting, wang2012superconductive, duan2014pressure, errea2020quantum, Alexey:CaB, Huan:Mg-Si, li2015pressure, liu2017potential, peng2017hydrogen, kvashnin2018high, zhang2022design, boeri20212021, yazdani2022artificial}, we believe that extra care are needed for the predictions. Readers, who are interested in the validity and the possible breakdown of the Migdal-\'{E}liashberg theory, are referred to some recent beautiful reviews \cite{chubukov2020eliashberg, marsiglio2020eliashberg} and original articles \cite{esterlis2018breakdown, alexandrov2001breakdown, hague2008breakdown, benedetti1998holstein, yuzbashyan2022breakdown}.

\subsection{First-principles computations}\label{sec:comput}
Computing the EP matrix element $g^{\nu}_{\bf k, k'}$ from first principles, and thus $\alpha^2F({\bf k}, {\bf k}', \omega)$ and $\alpha^2F(\omega)$, is crucial for using the Migdal-\'{E}liashberg theory in practice.\cite{pellegrini2024ab} The standard method for such computations is DFPT \cite{DFPT,giustino2017electron}, a perturbation treatment for the response of quantum systems described at the level of density functional theory (DFT) \cite{DFT1,DFT2}. Notable implementations of DFPT can be found in {\sc quantum espresso} \cite{QE1, QE2} and {\sc abinit} \cite{Gonze_Abinit_1,Gonze_Abinit_2,Gonze_Abinit_3}, two major DFT tools for calculating phonon-related properties of solids. Using these packages, thousands of predictions of superconductors have been reported since the 2000s \cite{pickett2022room, zurek2019high,sun2023clathrate, hilleke2022tuning, gao2021superconducting, zhao2023superconducting, boeri2019road,pickard2020superconducting, wang2012superconductive, duan2014pressure, errea2020quantum, Alexey:CaB, Huan:Mg-Si, li2015pressure, liu2017potential, peng2017hydrogen, kvashnin2018high, zhang2022design, boeri20212021, yazdani2022artificial, ribeiro2004possible, saniz2004electronic, di2021bh}. While the calculations of $\alpha^2F({\bf k}, {\bf k}', \omega)$ and $\alpha^2F(\omega)$ are computationally demanding, some high-throughput efforts \cite{shipley2021high, choudhary2022designing, wines2023high, cerqueira2024sampling, hai2023superconductivity} have made significant strides in this area.

In practice, $\alpha^2F({\bf k}, {\bf k}', \omega)$ and its isotropic version $\alpha^2F(\omega)$ are computed on two finite-size $\Gamma$-centered grids, one for the (electronic) $\bf k$ points, and one for the (phononic) ${\bf q}$ points defined as ${\bf q} = {\bf k} - {\bf k'}$, while the ${\bf q}$ grid must be a subgrid of the {\bf k} grid. Once $\alpha^2F({\bf k}, {\bf k}', \omega)$ and $\alpha^2F(\omega)$ are computed, there are two ways to estimate $T_{\rm c}$, as discussed in Sec. \ref{sec:theory}. In the first approach, Eqs. (\ref{Eq:El1}) and (\ref{Eq:El2}) are solved iteratively, e.g., on the imaginary axis, for a non-trivial solution of $\Delta({\bf k}, i\omega_n)$ using a designated code like Electron-Phonon Wannier ({\sc epw}) \cite{giustino2007electron,margine2013anisotropic, ponce2016epw}. Then, the real-axis superconducting gap function $\Delta_0(T)$ is approximated from $\Delta({\bf k}, i\omega_n)$, e.g., using P\'{a}de continuation \cite{marsiglio1988iterative}. Finally, $T_{\rm c}$ is determined as the maximum temperature at which the order parameter $\Delta_0(T)$ remains non-zero. The second, and more commonly employed, method uses empirical formulae for $T_{\rm c}$ from prior solutions to the \'{E}liashberg equations, such as the McMillan formula (\ref{ADM0}). Having $\alpha^2F(\omega)$, $\lambda$ and $\omega_{\rm log}$ can be computed using some post-processing tools \cite{marini2024epiq} of {\sc quantum espresso} suite \cite{QE1, QE2} and {\sc abinit} \cite{Gonze_Abinit_1,Gonze_Abinit_2,Gonze_Abinit_3} for Eqs. (\ref{eq:lambda}) and (\ref{eq:omlog}). Unfortunately, both approaches need an empirical value of $\mu^*$ for some reasons \cite{ponce2016epw, QE1, QE2, Gonze_Abinit_1,Gonze_Abinit_2,Gonze_Abinit_3}, while efforts to derive more rational values of $\mu^*$, e.g., using Eq. (\ref{eq:mu1}), are limited \cite{margine2016electron}.

Although the DFT-based calculations may be exact in principle, practical calculations of superconductivity-related quantities require certain finite-size $\bf k$- and $\bf q$-point grids, pseudopotentials, exchange-correlation functionals, finite-energy cutoffs, finite smearing widths for the $\delta$ functions in Eqs. (\ref{alpha2f_aniso}) and (\ref{eq:a2f}), empirical values of $\mu^*$, whether to include anharmonic effects\cite{troyan2021anomalous} and/or spin-orbit couplings\cite{kuderowicz2023strong} or not, and more \cite{errea2015high, akashi2015first,tran2023machine}. In many cases, the desired convergence, e.g., wrt the $\bf k$- and $\bf q$-point grids, may be computationally prohibitive, and some affordable parameters must be assumed. In fact, controlling these factors is challenging in phonon-related calculations, because acceptable numerical errors in energy and force evaluations are often orders of magnitude smaller than those deemed sufficient for standard DFT calculations. Despite the established reliability of DFT in regular electronic-structure problems \cite{lejaeghere2016reproducibility}, the above factors present potential sources of numerical error when computing $\alpha^2F({\bf k}, {\bf k}', \omega)$, $\alpha^2F(\omega)$, and ultimately, $T_{\rm c}$. A more detailed discussion on these challenges is presented in Sec. \ref{sec:challenge_theory}.

\begin{figure}[t]
\centering
\includegraphics[width=0.950\linewidth]{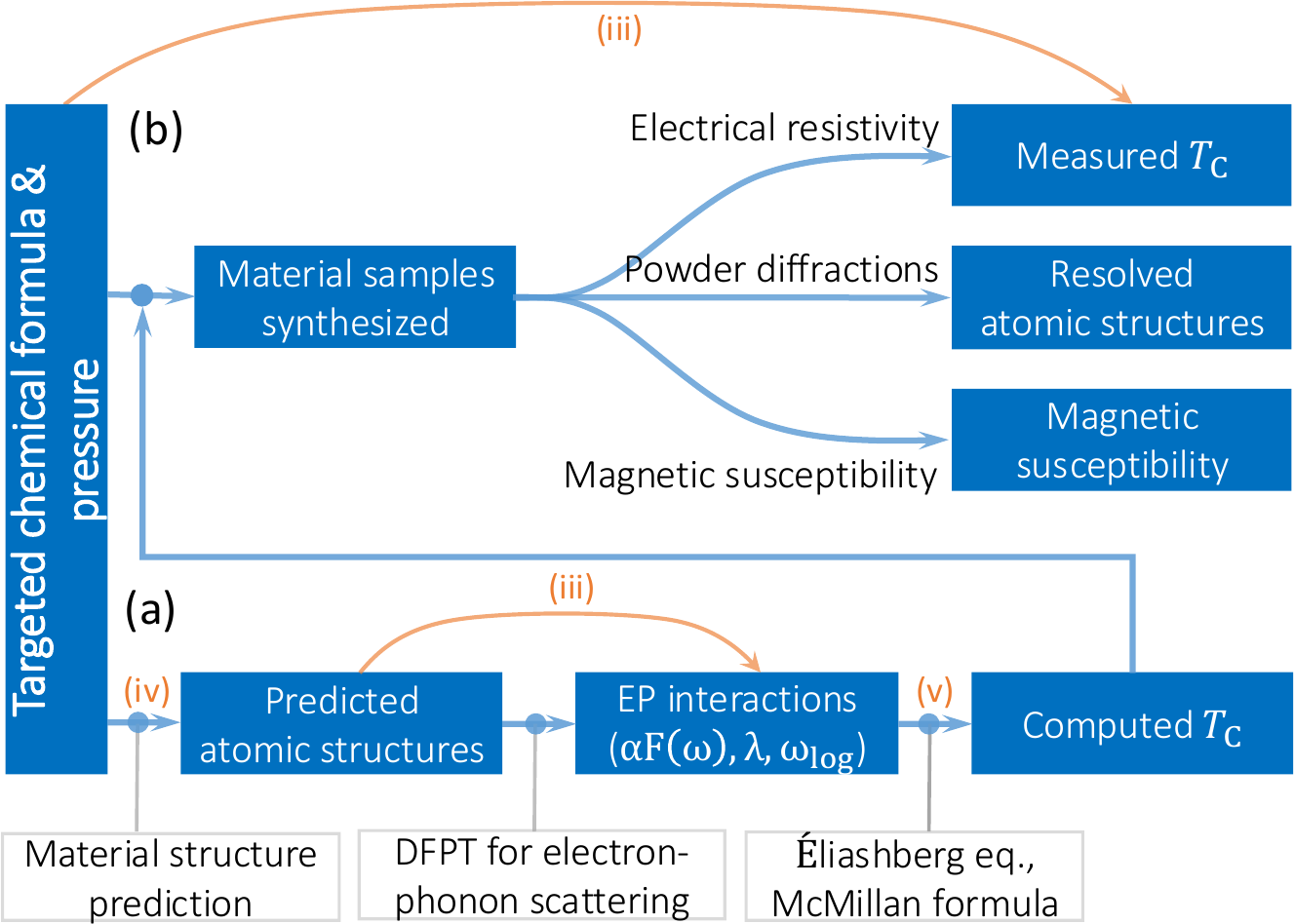}
\caption{Typical computational (a) and experimental (b) workflows for superconductor discovery. ML efforts are categorized in 5 groups (see Sec. \ref{sec:ML}), three of them are (iii) predicting superconductivity-related properties, (iv) accelerate the structure predictions by ML potentials, and (v) deriving new formulae of $T_{\rm c}$. Some computational predictions were advanced to experimental synthesis and characterization.}\label{fig:flow}
\end{figure}

\subsection{A practical computational discovery workflow}\label{sec:invdesign}
Targeted materials discovery, often referred to as ``inverse design'', involves the guided exploration of the vast materials space to find synthesizable materials possessing desired properties \cite{franceschetti1999inverse, sanchez2018inverse, gomez2018automatic, weymuth2014inverse, dAvezac:GA, Xiang:Sidesign}. In a broad sense, the materials space is truly infinite, containing all possible chemical compositions, external conditions, microscopic details, macroscopic morphologies, additives, and processing parameters needed to describe and fabricate a material. Due to this immense complexity and space, a {\it trial-and-error} approach is impractical, even with highly efficient evaluation and decision-making processes. Instead, an effective inverse-design strategy must involve an ``intelligent'' protocol that rationally directs the exploration toward the targeted properties. Limited versions of this protocol were developed \cite{franceschetti1999inverse, sanchez2018inverse, gomez2018automatic, weymuth2014inverse, dAvezac:GA, Xiang:Sidesign} actively contributing to the recent discovery of various functional materials \cite{lombardo2021artificial, tabor2018accelerating, tran2024polymer, hu2023recent, liu2023machine, Arun:design, Chiho:GA, Rohit:VAE}. One strategy, which is widely used for discovering superconductors \cite{pickett2022room, zurek2019high, sun2023clathrate, hilleke2022tuning, gao2021superconducting, zhao2023superconducting, boeri2019road,pickard2020superconducting, wang2012superconductive, duan2014pressure, errea2020quantum, Alexey:CaB, Huan:Mg-Si, li2015pressure, liu2017potential, peng2017hydrogen, kvashnin2018high, zhang2022design, boeri20212021, yazdani2022artificial, ribeiro2004possible, saniz2004electronic, di2021bh}, is depicted in Fig. \ref{fig:flow} (a). In this strategy, computational structure prediction methods for exploring the atomic configuration space reached an advanced level of maturity \cite{OganovBook,oganov2019structure, NeedsReview:CMD}.

The computational workflow for discovering new superconductors, as shown in Fig. \ref{fig:flow} (a), consists of several steps. First, usually guided by chemical intuition and expertise, a targeted chemical formula and external conditions (e.g., a range of pressures) are selected. Subsequently, employing methods like DFT, the atomic structures that are both thermodynamically and dynamically stable are predicted. In this step, a pressure-temperature phase diagram of the targeted chemical formula are usually needed.\cite{Huan:hafnia,Huan:Mg-Si,TuocSnS} For this purpose, one may need very expensive computations for the Gibbs free energy of different atomic phases, e.g., using advanced tools like {\sc phonopy}\cite{phonopy} and {\sc sscha}\cite{monacelli2021stochastic}. Next, the superconducting properties of these atomic structures are computed, by either solving the \'{E}liashberg equations or using the McMillan formula, to identify those with respectable predicted $T_{\rm c}$. While most of the computational works conclude at this juncture, some proceed to experimental synthesis and testing of these new materials, confirming the predicted superconductivity, as sketched in Fig. \ref{fig:flow} (b).

Putting the computations of phonon-mediated superconducting properties aside, atomic structure prediction is a very time consuming step. The goal is to find a thermodynamically stable arrangement of atoms for a given chemical formula at a given pressure, considering its potential decomposition into all possible related materials. This requires, at least, a convex hull analysis \cite{li2015pressure, liu2017potential, peng2017hydrogen, kvashnin2018high, semenok2021superconductivity, Alexey:CaB, Huan:Mg-Si, TuocSnS}, which necessitates the prediction of the lowest-enthalpy atomic structure, i.e., the global minimum of the multi-dimension potential energy surface (PES), for each related formula. Given the exponential increase in the number of local minima on a PES with the system size \cite{stillinger1999exponential}, the global optimization problem for each formula presents a formidable challenge. Despite its computational costs, an advanced first-principles based materials structure search is highly desired to explore {\it unknown domains} of the materials space in which {\it unknown chemistries and atomic structures} are expected, and relying on the known prototype structures is insufficient. In fact, unconstrained structure prediction endeavors have uncovered numerous novel (non-existing) atomic structures \cite{amsler12, Huan:Alanates, Huan:hafnia, Huan:perovskites} that were then confirmed experimentally \cite{Huang:LiAlH4Expr, wei2018rhombohedral, qi2020stabilization}. For superconductors discovery, the structure prediction step is critical, especially when the searches extend to the territories no one has ever explored, e.g., at hundreds of GPa. Current {\it state-of-the-art} structure prediction methods, which were used widely in superconductors discovery, are {\sc uspex} \cite{USPEX, USPEX:2}, {\sc calypso} \cite{CALYPSO}, {\it ab initio} random structure search (AIRSS) \cite{AIRSS}, XtalOpt \cite{XtalOpt}, {\sc maise} \cite{hajinazar2021maise}, and minima-hopping \cite{Goedecker:MHM, Amsler:MHM}.

\subsection{Experimental synthesis and characterizations}\label{sec:expt}
Some computationally discovered hydride superconductors were synthesized and tested experimentally \cite{eremets2008superconductivity, Drozdov15, somayazulu2019evidence, drozdov2019superconductivity,semenok2020superconductivity, chen2021synthesis, troyan2021anomalous, kong2021superconductivity, chen2021high,ma2022high, li2022superconductivity,song2023stoichiometric}. For those advanced to this step, the targeted formulations are typically synthesized by reacting pure metals with excess hydrogen or hydrogen-rich gases, e.g., ammonia borane and hydrocarbons, often using laser heating in diamond anvil cells (DAC) under the desired pressure.

\begin{figure}[t]
\centering
\includegraphics[width=0.95\linewidth]{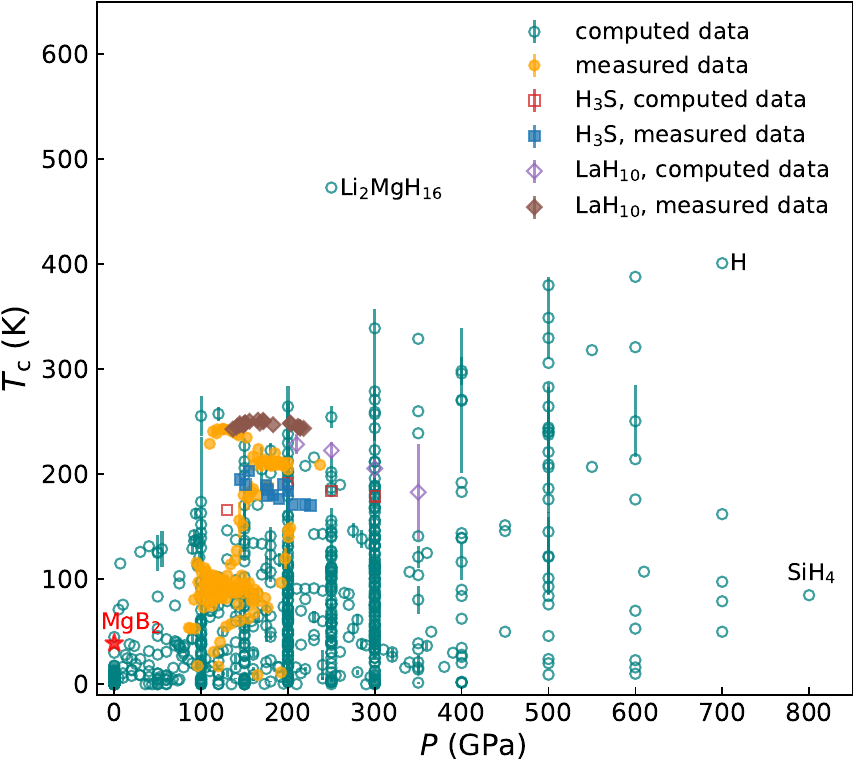}
\caption{An {\it incompleted} snapshot of the phonon-mediated superconductors discovered computationally. For some of them, e.g., H$_3$S and LaH$_{10}$, experimental data are available and also shown. Highlighted in the figure, data for MgB$_2$ at 0 GPa, Li$_2$MgH$_{16}$ at 250 GPa, H at 700 GPa, and SiH$_4$ at 800 GPa are taken from Refs. \citenum{nagamatsu2001superconductivity}, \citenum{sun2019route}, \citenum{zhong2022prediction}, and \citenum{zhang2015high}, respectively. Part of the data used for this Fig. has been used for Fig. 2c of Ref. \citenum{tran2023machine}.}\label{fig:disc}
\end{figure}

\begin{table*}[!htbp]
\caption{\footnotesize Notable hydride superconductors, given in terms of chemical formula, that were discovered computationally and then confirmed experimentally from the 2000s. For each of them, maximum observed critical temperature $T^{\rm max}_{\rm c}$, the pressure $P$ @ $T^{\rm max}_{\rm c}$ at which $T^{\rm max}_{\rm c}$ was observed, years and references of (experimental) discoveries and (computational) predictions, and a short summary of the discoveries, are provided.}\label{table:disc}
\vspace{-2mm}
\begin{footnotesize}
\begin{center}
\begin{tabular}{p{1.65cm} p{1.05cm} p{1.50cm} p{1.8cm} p{1.8cm} p{9.3cm}}
\hline
\multirow{2}{*}{Formula} & $T^{\rm max}_{\rm c}$ (K) & $P$ @ $T^{\rm max}_{\rm c}$ (GPa) & Exper. discovery& Comput. prediction & \multirow{2}{*}{Short summary}\\
\hline
SiH$_4$        & 17.5 & 90  &  2008 \cite{eremets2008superconductivity} & 2006, \cite{feng2006structures,pickard2006high} 2007 \cite{yao2007superconductivity} & Electrical resistance measured, $P6_3$ structure resolved, not matched with predictions, multiple structure searches follow, refining the structure\\
H$_3$S         & 203  & 155 & 2015 \cite{Drozdov15} & 2014 \cite{duan2014pressure} & Electrical resistance \& magnetic susceptibility measured, $Im\overline{3}m$ structure resolved \& matched with predictions, superconductivity replicated \cite{strobel2011novel, huang2019high, einaga2016crystal, nakao2019superconductivity, osmond2022clean}\\
LaH$_{10}$     & 250  & 170 &  2019 \cite{somayazulu2019evidence,drozdov2019superconductivity} & 2017 \cite{liu2017potential, peng2017hydrogen} & Electrical resistance measured, $Fm\overline{3}m$ structure resolved \& matched with predictions, superconductivity replicated \cite{struzhkin2020superconductivity}\\
ThH$_{9}$      & 146  & 170 & 2020 \cite{semenok2020superconductivity} & N/A & Electrical resistance measured, $P6_3/mmc$ structure resolved\\
ThH$_{10}$     & 161  & 174 & 2020 \cite{semenok2020superconductivity} & 2018 \cite{kvashnin2018high} & Electrical resistance measured, $Fm\overline{3}m$ structure resolved \& matched with predictions\\
PrH$_{9}$      & 8.4  & 120 & 2020 \cite{zhou2020superconducting} & 2020 \cite{zhou2020superconducting} & Electrical resistance measured, $P6_3/mmc$ structure resolved \& matched with predictions\\
BaH$_{12}$     & 20   & 140 &  2021 \cite{chen2021synthesis}& 2021 \cite{chen2021synthesis} & Electrical resistance measured, $Cmc2_1$ structure resolved \& matched with predictions\\
(La,Y)H$_{10}$ & 253  & 199 & 2021 \cite{semenok2021superconductivity}&2021 \cite{semenok2021superconductivity} & Electrical resistance measured, $Fm\overline{3}m$ structure resolved \& matched with predictions\\
YH$_6$         & 224     &  168   & 2021 \cite{troyan2021anomalous} & 2015 \cite{li2015pressure, peng2017hydrogen} & Electrical resistance measured, $Im\overline{3}m$ structure resolved \\
& 220     &  183   & 2021 \cite{kong2021superconductivity} & 2015 \cite{li2015pressure, peng2017hydrogen} & \& matched with predictions, superconductivity replicated \cite{kong2021superconductivity}\\
YH$_9$         & 243     & 201    & 2021 \cite{kong2021superconductivity} & 2017 \cite{peng2017hydrogen}& Electrical resistance measured, $P6_3/mmc$ structure resolved \& matched with predictions, superconductivity replicated \cite{wang2022synthesis}\\
CeH$_9$        &  57    &  88   &  2021 \cite{chen2021high} & 2017 \cite{peng2017hydrogen} & Electrical resistance measured, $P6_3/mmc$ structure resolved \& matched with predictions, $T_{\rm c}$ replicated \& diamagnetism of Meissner effect probed \cite{bhattacharyya2024imaging}\\
CeH$_{10}$     &  115    & 95    &  2021 \cite{chen2021high} & 2017 \cite{peng2017hydrogen} & Electrical resistance measured, $Fm\overline{3}m$ structure resolved \& matched with predictions\\
ScH$_3$        & 18.5 & 131 & 2021 \cite{shao2021superconducting}& 2010, \cite{kim2010general} 2016 \cite{wei2016pressure} & Electrical resistance measured, $Fm\overline{3}m$ structure resolved \& matched with predictions\\
LuH$_3$        & 12.4 & 122 & 2021 \cite{shao2021superconducting}& N/A & Electrical resistance measured, $Fm\overline{3}m$ structure resolved\\
CaH$_6$        & 215/210  & 172/160 &  2022 \cite{ma2022high, li2022superconductivity} & 2012 \cite{wang2012superconductive,shipley2021high} & Electrical resistance measured, $Im\overline{3}m$ structure resolved \& matched with predictions, superconductivity replicated \cite{li2022superconductivity}\\
SnH$_4$        & 72   & 180 &  2023 \cite{troyan2023non} & 2023 \cite{troyan2023non} & Electrical resistance measured, $Fm\overline{3}m$ structure resolved \& matched with predictions \\
LaBeH$_8$      & 110  & 80  &  2023 \cite{song2023stoichiometric} & & Electrical resistance measured, $Fm\overline{3}m$ structure resolved\\
NbH$_3$      & 42  & 187  &  2024 \cite{he2024superconductivity} & N/A& Electrical resistance measured, $Fm\overline{3}m$ structure resolved\\
La$_{0.5}$Ce$_{0.5}$ H$_{10}$  & \multirow{2}{*}{175}  & \multirow{2}{*}{155}  &  \multirow{2}{*}{2024 \cite{huang2024synthesis}} & \multirow{2}{*}{2024 \cite{huang2024synthesis}} & \multirow{2}{*}{Electrical resistance measured, $Fm\overline{3}m$ structure resolved}\\
Y$_{0.5}$Ce$_{0.5}$H$_{9}$  & 141  & 130  &  2024 \cite{chen2024synthesis} & N/A & Electrical resistance measured, $P6_3/mmc$ structure resolved\\
La$_4$H$_{23}$  & 90 & 95 &  2024 \cite{cross2024high} & NA & Electrical resistance measured, $Pm\overline{3}m$ structure resolved\\
    \hline
  \end{tabular}
\vspace{-5.5mm}
\end{center}
\end{footnotesize}
\end{table*}

Having the samples, the {\it electrical resistance} $R$ was always examined, typically using the four-point probe technique \cite{smits1958measurement, miccoli2015100th}. By definition, the presence of a superconducting transition is implied when $R$ drops sharply to zero at the critical temperature $T_{\rm c}$. The isotope effect \cite{de1954isotope, garland1963isotope}, a footprint of the lattice dynamics on the phonon-mediated superconductivity can then be observed in the dependence of $T_{\rm c}$ on the average isotope mass of the material. {\it Magnetic susceptibility measurements} are more challenging but critical \cite{goldfarb1991alternating, struzhkin2024magnetic} because they can help probing the Meissner effect \cite{meissner1933neuer}, a hallmark of superconductivity, while providing other essential information such as the critical field, penetration depth, and critical current density. Four methods that may be used are alternating-current (AC) susceptibility measurements using a pickup/compensating coil architecture \cite{huang2019high}, superconducting quantum interference device (SQUID) magnetometer \cite{Drozdov15}, synchrotron M\"{o}ssbauer technique \cite{troyan2016observation}, and a quantum sensing approach involving Nitrogen-vacancy centers implanted in DAC \cite{bhattacharyya2024imaging, ho2021recent}. As operating these techniques under hundreds of GPa in a DAC is non-trivial, only a few magnetic susceptibility measurements were reported \cite{huang2019high, struzhkin2020superconductivity}. Instead, the most widely-used approach is to measure the dependence of $T_{\rm c}$ on the external magnetic field, the behavior that could reveal the upper critical field, ultimately linking to the coherence length of the superconducting state \cite{landau1950theory,werthamer1966temperature}. A new method \cite{minkov2023magnetic}, measuring the magnetic flux trapped inside the superconducting samples, i.e., the incomplete Meissner effect, seems to be useful. Finally, {\it resolving the atomic structure} of the synthesized samples is desirable. For this goal, x-ray diffraction (XRD) pattern and sometimes Raman scattering are powerful methods. Overall, tremendous challenges remain for the experimental techniques in superconductor discovery, specifically magnetic susceptibility measurements at extremely high pressure in a DAC, and readers are referred to Refs. \citenum{bhattacharyya2024imaging, ho2021recent,minkov2023magnetic} for more information.

Fig. \ref{fig:flow} (b) summarizes the aforementioned measurements, among others, desired and used for probing the possible superconductivity. All the discovered hydride superconductors discussed in Sec. \ref{sec:disc} have been experimentally synthesized and at least one of the three characteristics are examined.

\section{Computation-driven superconductor discoveries}\label{sec:disc}
We now turn to the discoveries of hydride superconductors in the last two decades, mostly driven by the first-principles computational methods and workflow. It is interesting to note that Ashcroft's prediction \cite{ashcroft2004hydrogen, ashcroft2004bridgman} on the possible superconductivity in hydrides came more than three decades after his similar prediction on compressed hydrogen \cite{ashcroft1968metallic}. One way to view the difference between the two predictions is that hydrogen atoms in solid materials are ``chemically precompressed'' \cite{feng2006structures}, thus realizing the prediction in these materials could be technically more feasible. This may be a reason why the later \cite{ashcroft2004hydrogen, ashcroft2004bridgman} was soon followed by thousands of computational and dozens of experimental discoveries.

Table \ref{table:disc} and Fig. \ref{fig:disc} provide a (likely incompleted) summary of the computational and experimental discoveries reported from the 2000s. Not surprisingly, most of the materials are hydrogen-rich, predicted to be superconductors at very high pressures, ranging up to $\simeq 800$ GPa. One of the most striking computational predictions is the superconductivity of Li$_2$MgH$_{16}$ at 250 GPa with $T_{\rm c}\simeq 473$K, or 200$^\circ$C \cite{sun2019route}, while the first major experimental discovery is the superconductivity of H$_3$S at 155 GPa with $T_{\rm c} = 203$ K \cite{Drozdov15}. This discovery is remarkable because it reproduced a majority of the computational predictions reported one year earlier \cite{duan2014pressure}.

The remaining part of this Section is devoted to a detailed discussion on some superconducting materials that were predicted computationally and then synthesized and tested. Given the scope and the limited length of this Article, and an extensive list of experimental discoveries shown in Table \ref{table:disc}, we will examine a few of them that were reproduced independently and/or sparked significant interest and follow-up works from the community. For other discoveries, readers are referred to Table \ref{table:disc} in which a short summary and relevant references are given for each of them.

\begin{figure*}[t]
\centering
\includegraphics[width=0.7\linewidth]{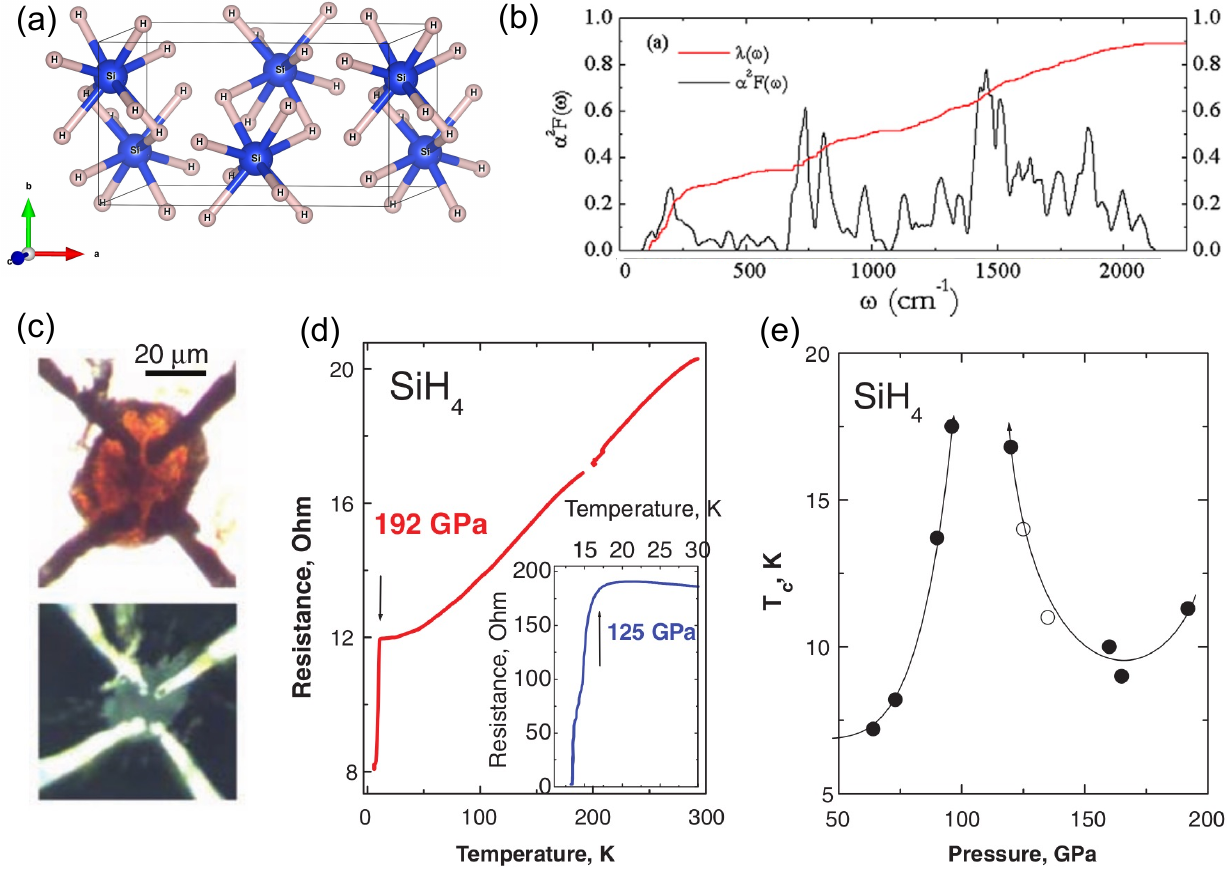}
\caption{(a) The predicted $C2/c$ structure of silane SiH$_4$ \citenum{yao2007superconductivity}, (b) the spectral function $\alpha^2F(\omega)$ and accumulated $\lambda (\omega)$ of the $C2/c$ structure, (c) the fabricated sample placed within the four-point probe, (d) a representative superconducting step in measured resistance, and (e) pressure-dependent $T_{\rm c}$ measured for silane. Panels (b) and (c, d, e) were reprinted with permission from Ref. \citenum{yao2007superconductivity} (Copyright 2007, IOP Publishing) and Ref. \citenum{eremets2008superconductivity} (Copyright 2008, The American Association for the Advancement of Science), respectively.}\label{fig:SiH4}
\end{figure*}

\subsection{Silane SiH$_4$}
Silane SiH$_4$ is the first hydride superconductor computationally predicted in 2006 \cite{feng2006structures, pickard2006high, yao2007superconductivity} and experimentally reported in 2008 \cite{eremets2008superconductivity}, shortly after Ashcroft's later prediction \cite{ashcroft2004hydrogen, ashcroft2004bridgman}. By examining multiple atomic structures of SiH$_4$ and computing the frequency cutoff $\omega_{\rm c}$ at a first-principles level, the orthohombic $Pman$ structure stands out with the predicted $T_{\rm c} \simeq 1.13\hbar\omega_{\rm c}/k_{\rm B}\exp(-1/N_{\rm F}V) \simeq 166$ K at 202 GPa \cite{feng2006structures}. In the same year, after a search for high-pressure structures of SiH$_4$, a qualitative assessment into the possible superconductivity, specifically involving a monoclinic $C2/c$ phase, was conducted with a preferable conclusion at pressures up to 50 GPa \cite{pickard2006high}. From 50 GPa to 250 GPa, an insulating tetragonal $I4_1/a$ phase was predicted \cite{pickard2006high} to be thermodynamically stable and then confirmed experimentally \cite{chen2008pressure, strobel2011high}. One year later, the computational workflow based on Migdal-\'{E}liashberg theory was used \cite{yao2007superconductivity} for another monoclinic $C2/c$ structure, visualized in Fig. \ref{fig:SiH4} (a), predicting a $T_{\rm c}\simeq 45-55$ K at 90 and 125 GPa. The spectral function $\alpha^2F(\omega)$ and the accumulated $\lambda (\omega)$ of this phase are shown in Fig. \ref{fig:SiH4} (b).

The computational predictions were followed by an experimental discovery \cite{eremets2008superconductivity}, in which SiH$_4$ samples, shown in Fig. \ref{fig:SiH4} (c), were fabricated. Electrical resistance measurements, one of them is given in Fig. \ref{fig:SiH4} (b), clearly show the superconducting-likes drop at a pressure-dependent critical temperature $T_{\rm c}$. Between $65$ GPa and $90$ GPa, $T_{\rm c}$ grows from 7K to 17.5 K before going down and up again in the regime of 120 -- 200 GPa, as shown in Fig. \ref{fig:SiH4} (e). This behavior suggests the involvement of more than one phases of SiH$_4$. Examining Raman scattering and XRD pattern, a superconducting hexagonal $P6_3$ phase was resolved in the first regime and the predicted $I4_1/a$ phase \cite{pickard2006high} was confirmed again in the second regime, in which both phases co-exist.

\begin{figure*}[t]
\centering
\includegraphics[width=0.70\linewidth]{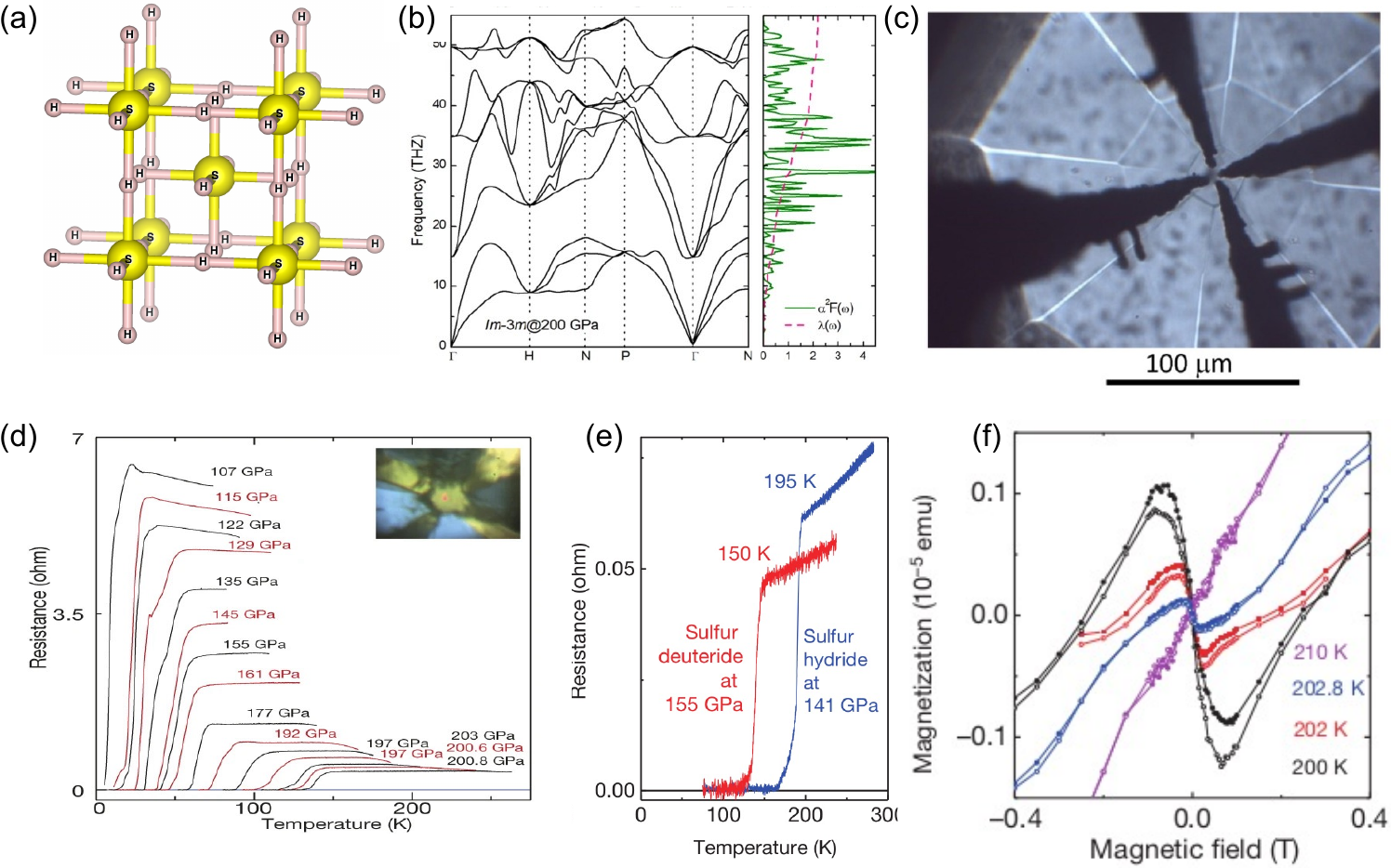}
\caption{(a) The predicted $Im\overline{3}m$ structure of H$_3$S, (b) phonon band structure, the spectral function $\alpha^2F(\omega)$, and accumulated $\lambda (\omega)$ of the $Im\overline{3}m$ structure at 200 GPa, (c) four Ti electrodes sputtered on a diamond anvil, i.e., a four-point probe, at the center of them is the sample, (d) $R$-$T$ dependence measured for H$_3$S at different pressures, (e) $R$-$T$ dependence measured for H$_3$S and H$_3$D, which reveals the isotope effect, and (f) the magnetization measured as functions of external field, showing the diamagnet and paramagnet characteristics below and above $T_{\rm c}\simeq 203$ K, respectively. Panels (b) and (c, d, e, f) were reprinted with permission from Ref. \citenum{duan2014pressure} (under a Creative Commons license) and Ref. \citenum{Drozdov15} (Copyright 2015, Springer Nature), respectively.}\label{fig:H3S}
\end{figure*}

While the superconducting hexagonal $P6_3$ phase was not predicted computationally, the discovery of the predicted superconductivity in SiH$_4$ and the verification of the predicted $I4_1/a$ phase are intriguing. In fact, they motivated multiple computational structure searches for SiH$_4$ \cite{chen2008superconducting, martinez2009novel, zhang2015high} and disilane, a related compound with formula Si$_2$H$_6$ \cite{jin2010superconducting, flores2012high}. As the $P6_3$ phase was found \cite{chen2008superconducting} to be dynamically unstable, a related orthorhombic $Cmca$ phase was suggested as a candidate for the superconducting phase of SiH$_4$. Among the structures subsequently found \cite{martinez2009novel} for SiH$_4$, an orthorhombic $Pbcn$ phase is not only thermodynamically more stable than the $Cmca$ phase but is also related to the $P6_3$ phase through its dynamical instability. Furthermore, computed $\lambda$ and $\omega_{\rm log}$ of the $Pbcn$ phase lead to $T_{\rm c} \simeq 16.5$ K by using the McMillan formula \cite{martinez2009novel}. To close this section, we refer readers to Ref. \citenum{strobel2011high} for a summary of such SiH$_4$-related efforts.

\subsection{Sulfur hydride H$_3$S}
In 2014, a structure prediction effort \cite{duan2014pressure} was performed to search for possible stable atomic structures of (H$_2$S)$_2$H$_2$ at elevated pressure $P$ up to 300 GPa. This work was partially motivated by an early experimental report \cite{strobel2011novel}, showing that H$_2$S and H$_2$, both are gas at ambient conditions, can form stoichiometric compound H$_3$S near 3.5 GPa. The computational examination \cite{duan2014pressure} reveals that when $P$ increases, H$_3$S compound undergoes a series of structural phase transitions from $P1$ to $Cccm$ at 37 GPa, then to $R3m$ at 111 GPa, and finally to $Im\overline{3}m$ at 180 GPa. By examining the simulated XRD pattern at 22 GPa, the stable $P1$ phase of H$_3$S was found \cite{duan2014pressure} to match with that experimentally reported \cite{strobel2011novel}. As the $R3m$ and $Im\overline{3}m$ phases are metallic at the pressures they are stable, superconducting-related calculations suggest that at 130 GPa, the $R3m$ phase has $\lambda = 2.07$, $\omega_{\rm log} = 1125.1$ K, which can be translated to $T_{\rm c}\simeq 155 - 166$ K using Eq. \ref{ADM0} with $\mu^*$ ranging from $0.10$ to $0.13$. Likewise, at 200 GPa, the $Im\overline{3}m$ phase has $\lambda = 2.19$, $\omega_{\rm log} = 1334.6$ K, and finally $T_{\rm c} \simeq$ 191 -- 204 K. The phonon band structure, the spectral function $\alpha^2F(\omega)$, and the accumulated $\lambda (\omega)$ of the $Im\overline{3}m$ phase of H$_3$S at 200 GPa are reproduced in Fig. \ref{fig:H3S} (a).

In 2015, the superconductivity with a record-breaking $T_{\rm c} = 203$ K at 155 GPa was observed on synthesized samples of H$_3$S (see Fig. \ref{fig:H3S} (c)) and reported \cite{Drozdov15}. Fig. \ref{fig:H3S} (d) shows the sharp drops of the electrical resistance to zero when $T$ is lowered to a critical value that depends on the pressure. A strong isotope effect was also observed \cite{Drozdov15} and shown in Fig. \ref{fig:H3S}(e), implying a link to the lattice phonons. Magnetic susceptibility measurements were performed using SQUID \cite{Drozdov15}, revealing a transition from the diamagnetic to the paramagnetic state at $\simeq 203.5$ K, as shown in Fig. \ref{fig:H3S}(f).

The superconductivity of superconductivity in H$_3$S was reproduced by some other groups \cite{huang2019high, einaga2016crystal, nakao2019superconductivity, osmond2022clean, troyan2016observation} In one effort \cite{huang2019high}, the superconductivity of H$_3$S was probed in a range of $P$ using AC magnetic susceptibility measurements. As $P$ increases, the superconductivity appears at 117 GPa with $T_{\rm c} = 38$ K. The critical temperature then rises to $183$ K at 149 GPa before lowering to 140 K at 171 GPa. This non-monotonic behavior may come from the stoichiometry change and the effects of $P$ on $\lambda$.

At first, there is possibility of the H$_2$S compound to involve in the observation \cite{Drozdov15}, but the predicted $T_{\rm c}\simeq 80$ K of H$_2$S at this pressure range \cite{Li:HS} is far from the measured data. Although the reported pressure (155 GPa) is in the regime where the $R3m$ of H$_3$S was predicted \cite{duan2014pressure}, the $Im\overline{3}m$ phase was anticipated \cite{Drozdov15} and then experimentally confirmed \cite{einaga2016crystal,nakao2019superconductivity, osmond2022clean, errea2015high}.  The values of $T_{\rm c}$ that were computed \cite{duan2014pressure} and measured \cite{Drozdov15} for H$_3$S at different pressure $P$ are summarized in Fig. \ref{fig:disc}, indicating an overall good agreement.

The discovery \cite{Drozdov15} and confirmation \cite{huang2019high, einaga2016crystal, nakao2019superconductivity, osmond2022clean} of the superconductivity in H$_3$S with strikingly high $T_{\rm c}$ came soon after the prediction \cite{duan2014pressure}. Moreover, the predicted superconducting $Im\overline{3}m$ phase was confirmed while the predicted $T_{\rm c}$ agrees well with the measured values. These factors make the superconductivity of H$_3$S a major discovery, sparking significant excitement not only in science community but also in media and society \cite{cho2015stinky}.

\begin{figure*}[t]
\centering
\includegraphics[width=0.70\linewidth]{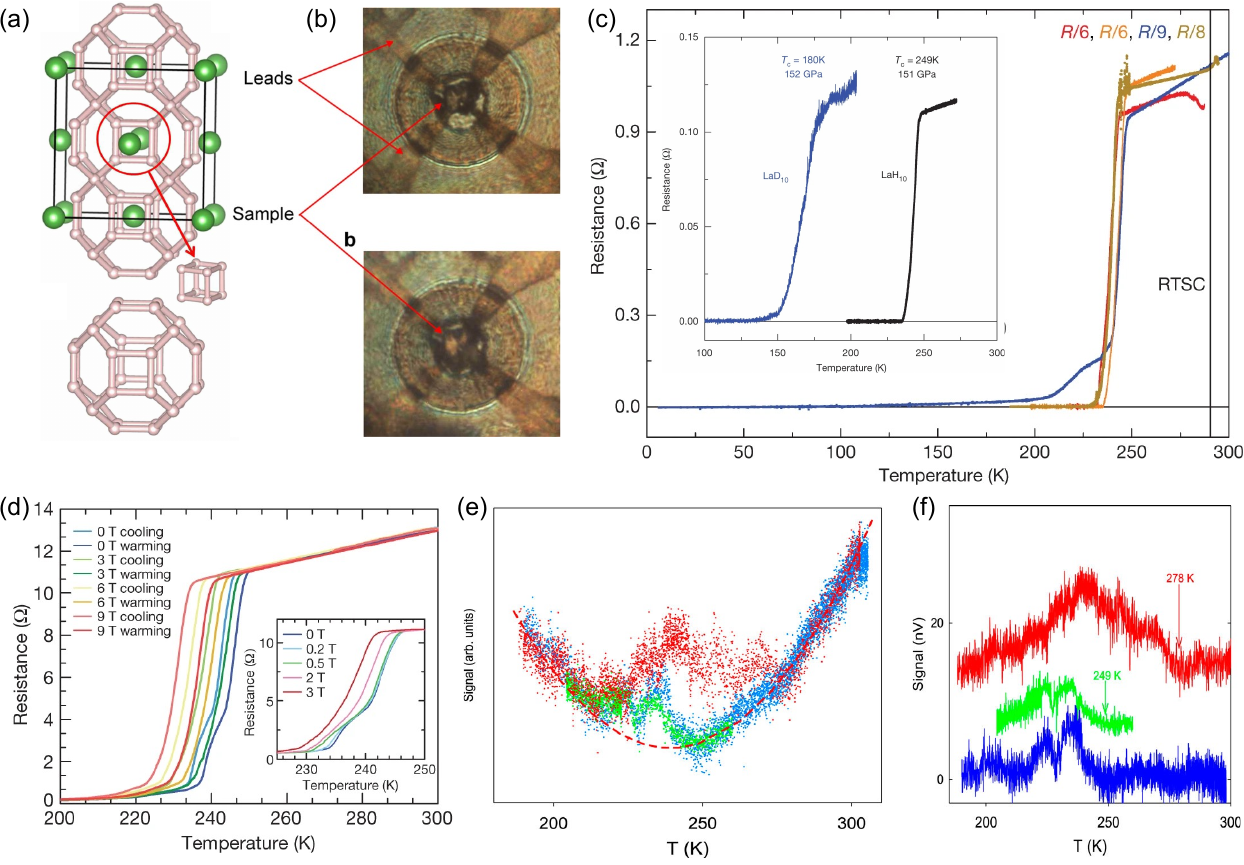}
\caption{(a) The $Fm\overline{3}m$ clathrate-like structure predicted for LaH$_{10}$, (b) LaH$_{10}$ samples fabricated and placed between four probes, (c) $R$-$T$ dependence measured for LaH$_{10}$ samples at zero external magnetic field, (d) $R$-$T$ dependence measured for LaH$_{10}$ samples at varying external magnetic field up to 9 T, (e) magnetic response signals from LaH$_{10}$ samples in a DAC, whose the background is shown in dahsed line, and (f) magnetic response after removing the background signal. The inset of panel (c) show the $R$-$T$ curves measured for LaH$_{10}$ and LaD$_{10}$, which reveal the isotope effect. Panels (a), (b, c, d), and (e, f) were reprinted with permission from Ref. \citenum{liu2017potential} (Copyright 2017, National Academy of Sciences of the United States of America), Ref. \citenum{drozdov2019superconductivity} (Copyright 2019, Springer Nature), and Ref. \citenum{struzhkin2020superconductivity} (under the Creative Commons CC BY license), respectively.}\label{fig:LaH10}
\end{figure*}

\subsection{Lanthanum hydride LaH$_{10}$}
The discovery of the superconductivity in lanthanum hydride LaH$_{10}$ was also initiated by computational structure prediction endeavors. In 2017, dozens of clathrate-like structures of hydrides were discovered computationally \cite{liu2017potential, peng2017hydrogen} at high pressures with very high predicted $T_{\rm c}$. In these prototype structures, rare-earth elements like La and Y are located at the center of hydrogen cavities, linked together throughout the space (see Figs. \ref{fig:LaH10}(a) and \ref{fig:YH69} (d) for visualizations). Among the examined rare-earth hydrides, LaH$_{10}$ in its cubic $Fm\overline{3}m$ clathrate-like structure was predicted to have an estimated $T_{\rm c} \simeq 288$K at 200 GPa \cite{peng2017hydrogen} and $T_{\rm c} \simeq 274-286$K at 210 GPa \cite{liu2017potential}. Quickly, LaH$_{10}$ was synthesized \cite{geballe2018synthesis} in the predicted $Fm\overline{3}m$ clathrate structure at 170 GPa, the pressure it was predicted computationally to be dynamically unstable \cite{liu2017potential}. The small discrepancy between computations and experiments was then attributed to lattice vibrations \cite{liu2018dynamics}, neglected in early computations \cite{liu2017potential}. Nevertheless, the experimental synthesis of LaH$_{10}$ in the predicted structure is a success.

The superconductivity of LaH$_{10}$ was observed in 2019 \cite{somayazulu2019evidence, drozdov2019superconductivity}. Superconducting-like drops in the measured electrical resistance $R$ of the synthesized samples were reported to be at $T_{\rm c} \simeq 260$ K under $180-200$ GPa \cite{somayazulu2019evidence} and $T_{\rm c} \simeq 250$ K under 170 GPa \cite{drozdov2019superconductivity}, as shown in Fig. \ref{fig:LaH10} (c). In Ref. \citenum{drozdov2019superconductivity}, the cubic $Fm\overline{3}m$ structure of LaH$_{10}$ was confirmed while other characteristics of superconductivity, including the isotope effect, shown in the inset of Fig. \ref{fig:LaH10} (c), and the reduce of $T_{\rm c}$ under increasing magnetic field, shown in Fig. \ref{fig:LaH10} (d), were also observed. By obtaining the upper critical field as a function of $T$ and fitting the data into the Ginzburg-Landau model \cite{landau1950theory}, a coherence length of about $1.56 - 1.86$ nm was extracted \cite{drozdov2019superconductivity}. However, magnetization measurements from the DAC cannot be performed using SQUID because of the small volume and size (10 -- 20 $\mu$m) of the samples.

One year later, the synthesis of LaH$_{10}$ was repeated \cite{struzhkin2020superconductivity}. The samples are not larger, but using the pickup/compensating coil technique, weak but measurable signals from magnetic susceptibility measurements, as shown in Fig. \ref{fig:LaH10}(e), were obtained \cite{struzhkin2020superconductivity}. After removing the background, the final data, shown in Fig. \ref{fig:LaH10}(f), point to superconducting transitions at $T_{\rm c} \simeq 250 - 280$ K and $P \simeq 170 - 180$ GPa \cite{struzhkin2020superconductivity}. In 2021, the cubic $Fm\overline{3}m$ phase was confirmed again while some other key superconducting characteristics such as the upper critical ﬁelds and coherence lengths was established \cite{sun2021high}.

The clathrate structures realized in LaH$_{10}$, in which the hydrogen cavities are interconnected and distributed continuously throughout the space (see Figs. \ref{fig:LaH10}(a)), allows for unusually high hydrogen content. For that reason, it can be viewed a close realization of metallic hydrogen, for which high-$T_{\rm c}$ values were predicted in 1968, also by Ashcroft \cite{ashcroft1968metallic}. Remarkably, this host-atoms-in-hollows motif has been predicted computationally \cite{wang2012superconductive, li2015pressure, peng2017hydrogen, sun2023clathrate} and then realized experimentally \cite{troyan2021anomalous, kong2021superconductivity} for many superconducting hydrides such as YH$_{6}$, YH$_{9}$, CeH$_{9}$ and CeH$_{10}$, as discussed later. Some computational works follow this promising lead, suggesting even hydrogen-richer clathrate structures of superhydrides, e.g., LaH$_{16}$ \cite{kruglov2020superconductivity}, LaH$_{18}$, YH$_{18}$, AcH$_{18}$, CeH$_{18}$, and ThH$_{18}$ \cite{zhong2022prediction, song2022potential} with predicted $T_{\rm c} \gtrsim 200$ K.

\subsection{Yttrium hydrides YH$_{6}$ and YH$_{9}$}\label{sec:Y}
Yttrium hydrides YH$_{6}$ and YH$_{9}$ are in the list of ten high-$T_{\rm c}$ superconducting rare-earth hydrides, including LaH$_{10}$, computationally predicted in 2017 \cite{peng2017hydrogen}. The common feature of this series is their clathrate-like structure, which, as mentioned above, is interesting in the context of high-$T_{\rm c}$ superconductivity \cite{sun2023clathrate}. The predicted $T_{\rm c}$ of YH$_{6}$ (in its cubic $Im\overline{3}m$ phase) and YH$_{9}$ (in its hexagonal $P6_3/mmc$ phase) are $T_{\rm c} \simeq 250$ K at 120 GPa and $T_{\rm c} \simeq 260$ K at 150 GPa, respectively \cite{peng2017hydrogen}. In fact, the cubic $Im\overline{3}m$ phase of YH$_{6}$ was also predicted to be a superconductor with $T_{\rm c} \simeq 251-264$ K at 164 GPa in 2015 in a structure prediction campaign \cite{li2015pressure}.

\begin{figure*}[t]
\centering
\includegraphics[width=0.7\linewidth]{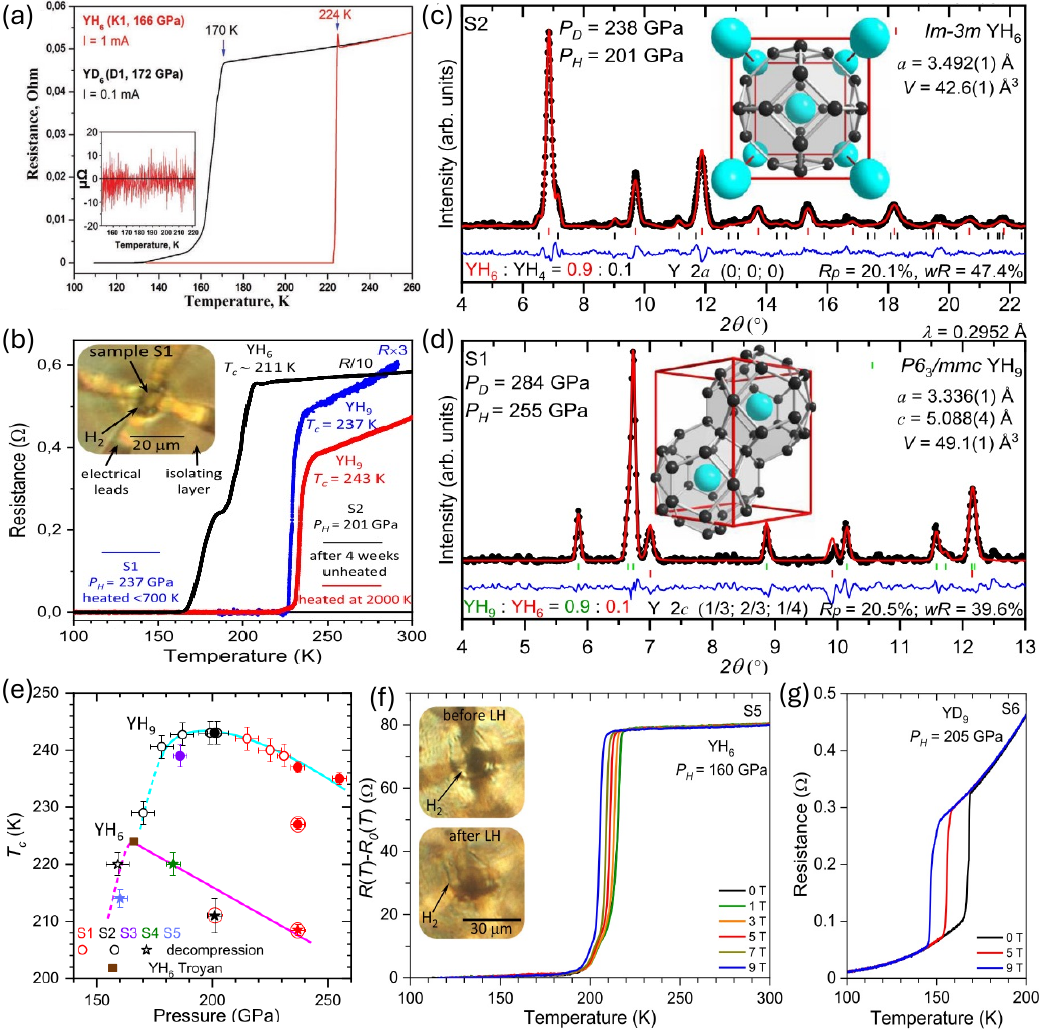}
\caption{(a) $R$-$T$ dependence measured for YH$_6$ (166 GPa) and YD$_6$ (172 GPa), (b) $R$-$T$ dependence measured for YH$_6$ (237 GPa) and YH$_9$ (201 GPa), (c) XRD-resolved $Im\overline{3}m$ clathrate-like structure of YH$_6$, (d) XRD-resolved $P6_3/mmc$ clathrate-like structure of YH$_9$, (e) $P$-dependent $T_{\rm c}$ measured for YH$_{6}$ and YH$_{9}$, and $R$-$T$ dependence measured for YH$_6$ (f) and YH$_9$ (g) at varying external magnetic field. Panels (a) and (b, c, d, e, f, g) were reprinted with permission from Ref. \citenum{troyan2021anomalous} (Copyright 2021, John Wiley and Sons) and Ref. \citenum{kong2021superconductivity} (under the Creative Commons CC BY license), respectively.}\label{fig:YH69}
\end{figure*}

The experimentally observed superconductivity of YH$_{6}$ and YH$_{9}$ was reported in Refs. \citenum{troyan2021anomalous} and \citenum{kong2021superconductivity}, both in 2021. First, YH$_6$ was synthesized and observed in the predicted $Im\overline{3}m$ phase \cite{troyan2021anomalous}. The measured electrical resistance $R$, shown in Fig. \ref{fig:YH69} (a), has a superconducting-like drop at $T_{\rm c} \simeq 224$ K at 168 GPa. The isotope effect was also observed, suggesting a link between the superconductivity and the lattice phonons. This discovery was reproduced in Ref. \citenum{kong2021superconductivity} where a $T_{\rm c} \simeq 220$ K at 183 GPa was reported for the $Im\overline{3}m$ phase of YH$_{6}$, as shown in Figs. \ref{fig:YH69} (b), \ref{fig:YH69} (c), and \ref{fig:YH69} (e). The superconductivity of YH$_{9}$ in its predicted hexagonal $P6_3/mmc$ phase (see Fig. \ref{fig:YH69} (d)) was also reported in Ref. \citenum{kong2021superconductivity}, exhibiting a pressure-dependent $T_{\rm c}$ peaking $\simeq 243$ K at 201 GPa (see Fig. \ref{fig:YH69} (e)). The dependence of $T_{\rm c}$ on the external magnetic field is shown in Fig. \ref{fig:YH69} (f) and Fig. \ref{fig:YH69} (g), pointing to an upper critical field and a coherence length. Fitting the measured $T$-dependent upper critical fields to the Ginzburg-Landaul \cite{landau1950theory} and Werthamer-Helfand-Hohenberg \cite{werthamer1966temperature} models, the coherence length at 0 K of YH$_{6}$ and YH$_{9}$ is $1.45-1.75$ nm and $2.3-2.7$ nm, respectively.

The superconducting-like drops in the electrical resistance $R$, the dependence of $T_{\rm c}$ on external magnetic field, and the atomic structure of both YH$_{6}$ and YH$_{9}$ were reproduced recently \cite{wang2022synthesis}. Nevertheless, magnetic susceptibility measurements have not been reported thus far, perhaps because of known technical challenges. In this context, emerging techniques to visualize local domain of diamagnetism \cite{bhattacharyya2024imaging} or measuring the trapped flux \cite{minkov2023magnetic} could be useful, as demonstrated \cite{bhattacharyya2024imaging} for CeH$_{9}$ and mentioned in Sec. \ref{sec:CeH910}.

\subsection{Cerium hydrides CeH$_{9}$ and CeH$_{10}$}\label{sec:CeH910}
\begin{figure*}[t]
\centering
\includegraphics[width=0.70\linewidth]{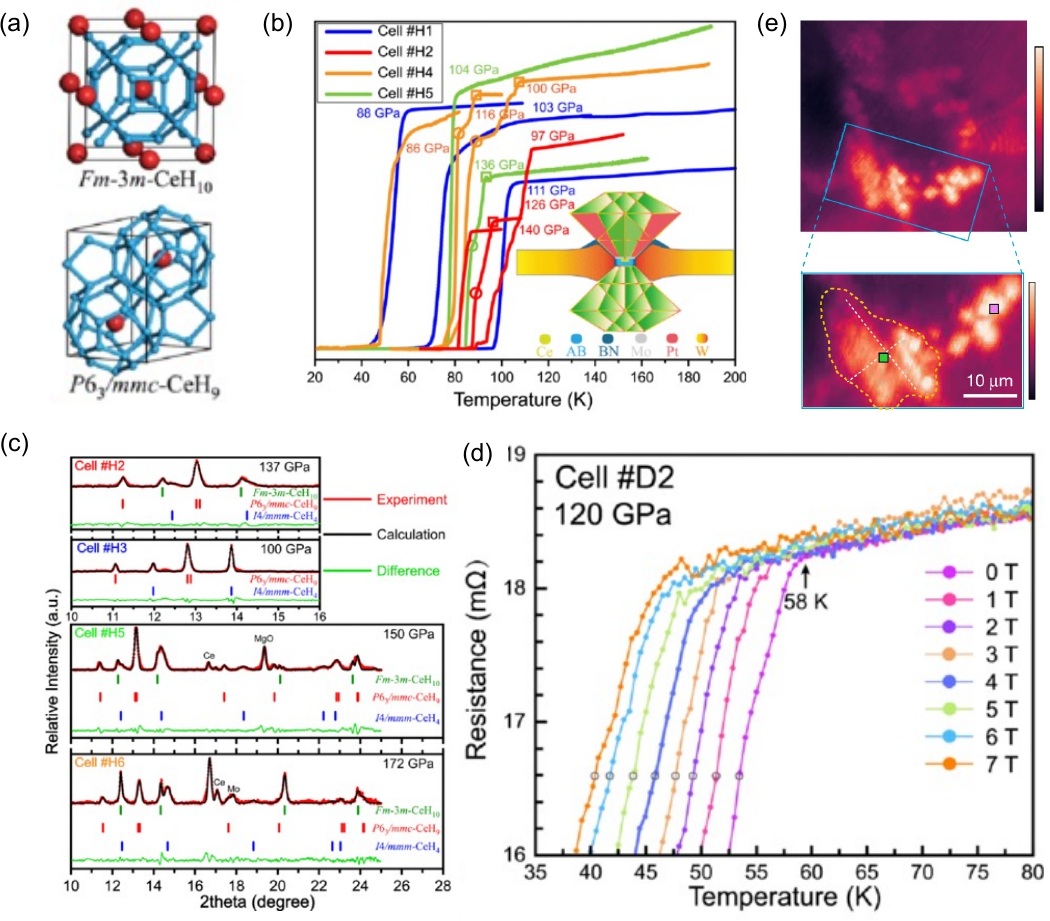}
\caption{(a) Predicted $Fm\overline{3}m$ structure of CeH$_{10}$ and $P6_3/mmc$ structure of CeH$_{9}$, (b) $R$-$T$ dependence measured for CeH$_9$ and CeH$_{10}$, (c) XRD-resolved $Fm\overline{3}m$ structure of CeH$_{10}$ and $P6_3/mmc$ structure of CeH$_{9}$, (d) $R$-$T$ dependence measured for CeH$_9$ and CeH$_{10}$ at varying external magnetic field, and (e) confocal fluorescence image of CeH$_{9}$ in which the diamagnet domains are shown in bright color. Panels (a, b, c, d) and (e) were reprinted with permission from Ref. \citenum{chen2021high} (Copyright 2021, American Physical Society) and Ref. \citenum{bhattacharyya2024imaging} (Copyright 2024, Springer Nature), respectively.}\label{fig:CeH910}
\end{figure*}

Similar to LaH$_{10}, $YH$_{6}$, and YH$_{9}$, CeH$_{9}$ and CeH$_{10}$ are also in the series of rare-earth hydrides predicted in Ref. \citenum{peng2017hydrogen} The predicted structures of CeH$_{9}$ and CeH$_{10}$, visualized in Fig. \ref{fig:CeH910}(a), are both clathrate-like, which belong to the $P6_3/mmc$ and $Fm\overline{3}m$ space groups, respectively. Although the predicted $T_{\rm c}$ of CeH$_{9}$ and CeH$_{10}$ is lower than others, i.e., $\simeq 50-60$ K, the pressure at which the superconductivity was predicted is relatively lower, i.e., 100 GPa for CeH$_{9}$ and 200 GPa for CeH$_{10}$ \cite{peng2017hydrogen}. Given that the superconductivity of H$_3$S, LaH$_{10}$, YH$_{6}$, and YH$_{9}$ was reported at close to 200 GPa, this prediction stimulates some interest. Amid the search, CeH$_{9}$ was the synthesized \cite{salke2019synthesis}, confirming the predicted $P6_3/mmc$ phase. A structure prediction campaign was then launched \cite{salke2019synthesis}, recovering the predicted $P6_3/mmc$ and $Fm\overline{3}m$ phases of CeH$_{9}$ and CeH$_{10}$, respectively. Calculations performed return $\lambda = 2.3$ and $\omega_{\rm log} = 740$ K for the $P6_3/mmc$ phase of CeH$_{9}$ at 200 GPa, which were translated to $T_{\rm c}$ = $105-117$ K using the McMillan formula with $\mu^* = 0.10-0.13$ \cite{salke2019synthesis}.

In 2021, the superconductivity of CeH$_{9}$ and CeH$_{10}$ was reported at 88 GPa and 95 GPa, respectively \cite{chen2021high}. The critical temperature, extracted from measured electrical resistance $R$ (shown in Fig. \ref{fig:CeH910}(b)), is $57$ K for CeH$_{9}$ and $115$ K for CeH$_{10}$ \cite{chen2021high}. The $P6_3/mmc$ phase of CeH$_9$ is predominant in the synthesized samples while the CeH$_{10}$ in its $Fm\overline{3}m$ phase plays a minor role, as shown in Fig. \ref{fig:CeH910}(c). Using the Ginzburg-Landau \cite{landau1950theory} and Werthamer-Helfand-Hohenberg \cite{werthamer1966temperature} models for the magnetic field-dependent $T_{\rm c}$ measured data (see Fig. \ref{fig:CeH910}(d)), the upper critical field at 0 K of CeH$_{9}$ was estimated to be is $17.7-22.9$ T at 140 GPa, while the coherence length is 3.4 nm at 120 GPa, 4.3 nm at 150 GPa, and 5.1 nm at 150 GPa \cite{chen2021high}. An isotope effect was observed for CeH$_{9}$ \cite{chen2021high}, but no magnetic susceptibility measurements were reported.

Recently, local diamagnetic domains of CeH$_{9}$ were observed \cite{bhattacharyya2024imaging} using a new technique that is capable of performing local magnetometry with sub-micron spatial resolution inside a DAC \cite{bhattacharyya2024imaging}. These domains, shown in bright colors in Fig. \ref{fig:CeH910}(e), are about 10 $\mu$m in size and small for SQUID. This result is intriguing, suggesting that the new technique would be useful for probing the superconductivity in small samples, the scenario that is common in superconductor discovery \cite{drozdov2019superconductivity}.

\section{Materials informatics in superconductor discovery}\label{sec:ML}
Emerging in the early 2010s and partly propelled by the Materials Genome Initiative \cite{MGI}, materials informatics has rapidly developed into a widely-used tool in materials research \cite{Rampi:ML, agrawal2016perspective, jain2016new, draxl2018nomad, draxl2019nomad, draxl2020big, Luca:descriptor, damewood2023representations}. This approach employs AI/ML techniques to learn materials data, creating models that can generate rapid predictions and complement traditional approaches in accelerating materials discovery. Synergetic approaches involving materials informatics methods, simulations, and physical experimentations have proactively driven numerous recent materials discoveries. Many of these newly discovered materials have been synthesized and tested. Notable examples include battery materials \cite{lombardo2021artificial}, green energy materials \cite{tabor2018accelerating}, functional and sustainable polymers \cite{tran2024polymer}, alloys \cite{hu2023recent, liu2023machine}, and more.

\begin{figure}[b]
\centering
\includegraphics[width=0.950\linewidth]{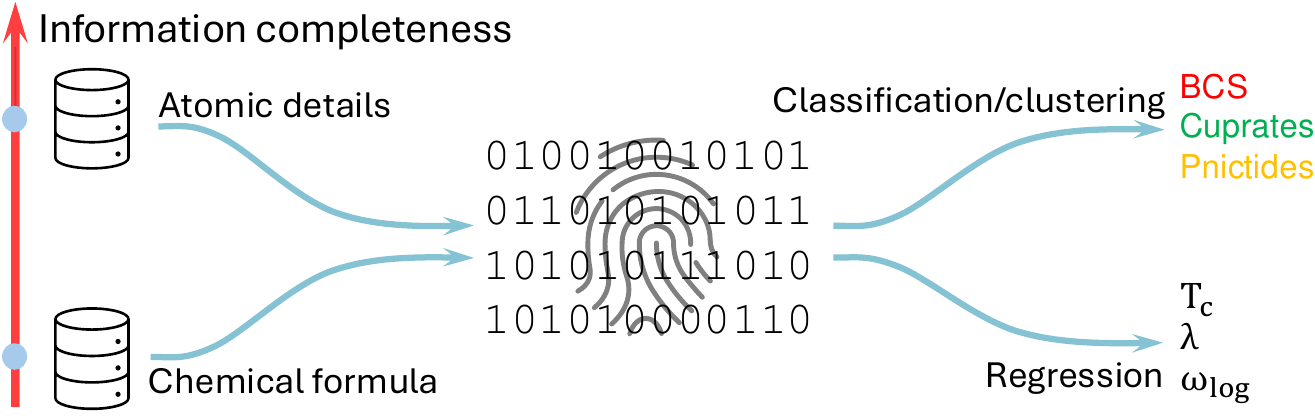}
\caption{A widely adopted schematic pipeline for ML predictive models of superconductivity.}\label{fig:ml1}
\end{figure}

Data is the cornerstone of materials informatics. In this field, data can originate from different sources such as experimental measurements and/or physics-based computations, be represented in different formats, contain different amounts and details of information, and possess different levels of fidelity. An ideal dataset for ML model training should be sufficiently big in size, diverse in the chemical, configuration, and parameter spaces, and complete in the relevant information. The development of major DFT-based materials databases like Materials Project \cite{Jain2013}, OQMD \cite{OQMD}, AFLOWLIB \cite{AFLOWLIB}, and NOMAD \cite{draxl2018nomad, draxl2019nomad} is important for future the progress of materials informatics.

For problems related to superconductivity, a typical model development procedure, sketched in Fig. \ref{fig:ml1}, starts with collecting and preparing training data for related materials. This data includes both the structural information of the materials and the target labels or properties of interest. In a next step, the materials, represented either by their chemical formulae or atomic structures, are converted into numerical ``hand-crafted'' features that can be readily interpreted and utilized by ML algorithms \cite{Luca:descriptor, damewood2023representations, Rampi:ML}. This step is required from the early days, but recent advances in deep-learning techniques \cite{agrawal2019deep,  lee2021transfer, batzner2022, kaba2022equivariant, gong2023general, zeng2019atom, konno2021deep, KuennethPolyBERT2023, gurnani2023polymer} signal that the materials features may also be ``learned''. Subsequently, a suitable ML algorithm learns the features, establishing a clustering, classification, or regression model capable of rapidly predicting or evaluating new materials.

As the model input, a chemical formula inherently provides less detailed information than a fully specified atomic structure. On the spectrum of information completeness, as previously discussed \cite{fan2012information} and sketched in Fig. \ref{fig:ml1}, the information content of a chemical formula is lower than that of an atomic structure. From the physics standpoint, each formula corresponds to an infinite number of metastable atomic structures (local minima on the PES) \cite{stillinger1999exponential}. Many of these structures are energetically close to the global minimum, implying that they can be stabilized in experiments or appear in computational models. Importantly, these low-energy structures have completely different atomic arrangements and properties, with one being insulating while another is conducting \cite{HuanData2021}. Superconductivity, as discussed in Sec. \ref{sec:theory}, is highly sensitive to the atomic arrangements. Thus, describing a superconductor only by its chemical formula entails the risk of inaccurate results. Eventually, models relying only on formulae as inputs inevitably encounter a degree of irreducible (aleatoric) uncertainty in their predictions that cannot be simply reduced by increasing the amount of data \cite{fan2012information, Tuoc:PrNN}. As early superconductor databases provide only chemical formulae, recent efforts to bring atomic details in, as discussed in Sec. \ref{sec:data}, have become a focal point of research.

Extending some previous discussions \cite{tran2023machine, bai2024unveiling}, materials informatics efforts in superconductor discovery can be categorized into five groups. The first group (i) includes those fully devoted to \cite{yamaguchi2020sc, yamaguchi2022superconductivity, foppiano2019proposal, foppiano2021supermat, foppiano2023automatic, sommer20233dsc} or partially involved in \cite{stanev2018machine, choudhary2022designing, hamidieh2018data, tran2023machine, cerqueira2024sampling} data generation, extraction, and dissemination. The other groups are (ii) classifying superconductors \cite{isayev2015materials,stanev2018machine, roter2022clustering}, (iii) predicting superconducting-related properties \cite{hamidieh2018data, matsumoto2019acceleration,pereti2023individual,ishikawa2019materials, hutcheon2020predicting, shipley2021high, zhang2022machine, stanev2018machine, konno2021deep, cerqueira2024sampling, choudhary2022designing,tran2023machine}, (iv) accelerating the structure prediction step with ML potentials \cite{gu2019superconducting, yang2021hard, dolui2024feasible}, and (v) refining empirical formulae for $T_{\rm c}$ \cite{shipley2021high, xie2019functional, xie2022machine}. Among these groups, (iii), (iv), and (v) are sketched in Fig. \ref{fig:flow}. Due to the integrated nature of the materials informatics approaches, the availability of the data, and specifically the forward-looking aspiration of the field, subsequent discussions in this Section will involve all classes of superconductors, including those mediated by phonons. The current status, challenges, and critical next steps of the materials informatics efforts in superconductor discovery will be discussed in Sec. \ref{sec:challenge}.

\subsection{Data generation, extraction, and dissemination}\label{sec:data}
The Superconducting Material Database maintained by Japan's National Institute for Materials Science (NIMS) provides data for many efforts in the field \cite{isayev2015materials, hamidieh2018data, stanev2018machine, roter2022clustering}. This database, referred to as \texttt{NIMS SuperCon}, records the chemical formula of about 32,000 known conventional (phonon-mediated) and unconventional superconductors, not all of them have (measured) $T_{\rm c}$ included. Recently, \texttt{NIMS SuperCon} was cleaned, re-edited, and released as \texttt{MDR SuperCon}, containing 26,323 records. Parallel cleaning and curation efforts \cite{hamidieh2018data, stanev2018machine} resulted in another version, presently known as \texttt{SuperCon}, containing about 16,400 records, among them 4,000 records have no $T_{\rm c}$. Details on \texttt{NIMS SuperCon}, \texttt{MDR SuperCon}, and \texttt{SuperCon} are given in Table \ref{table:data}.

\begin{figure*}[t]
\centering
\includegraphics[width=0.70\linewidth]{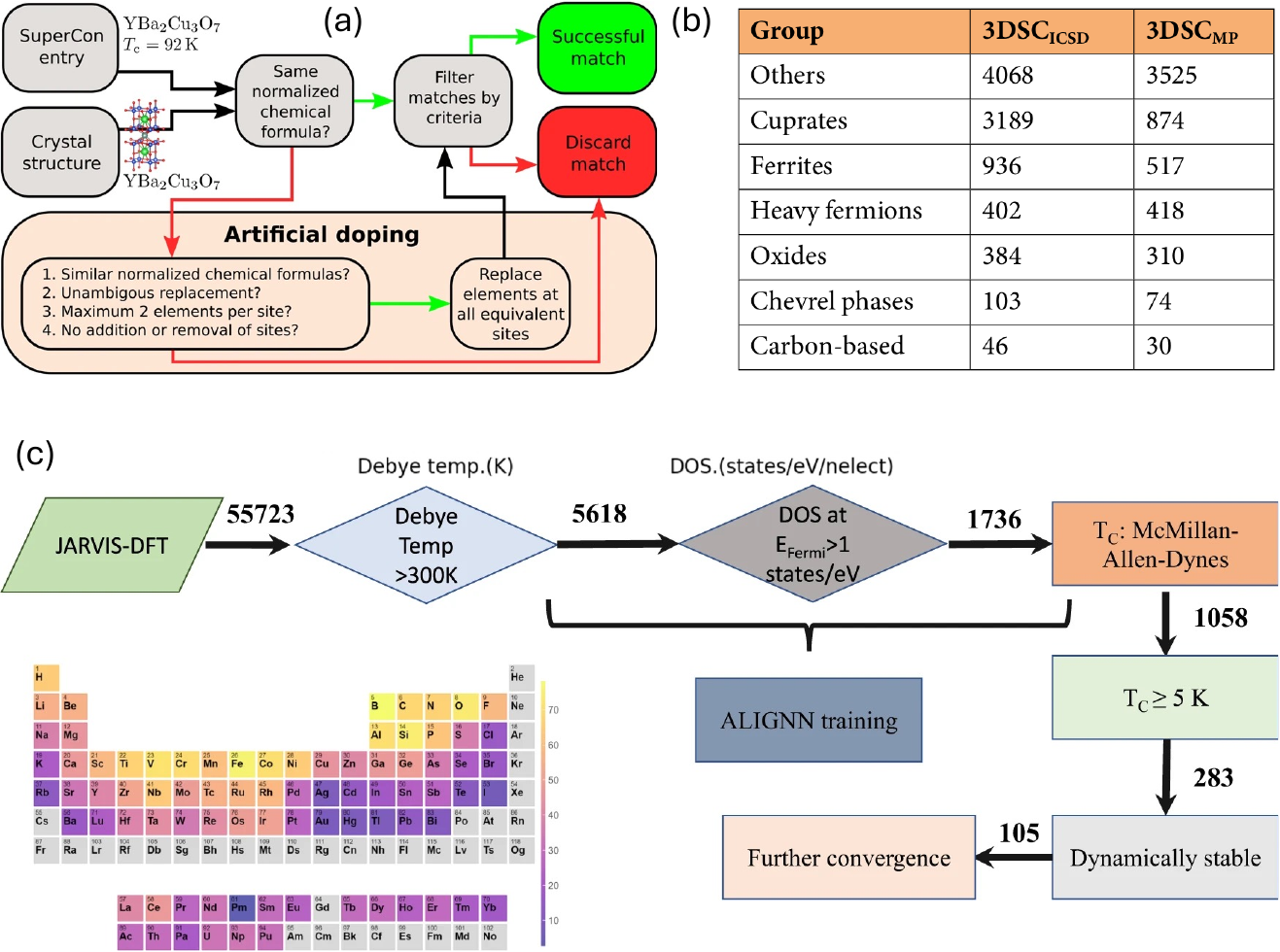}
\caption{Workflows used to develop (a) \texttt{3DSC}$_{\rm ICSD}$, \texttt{3DSC}$_{\rm MP}$ and (c) \texttt{Jarvis\_EPC}, three datasets of superconductivity-related parameters and atomic structures.  A summary of \texttt{3DSC}$_{\rm ICSD}$ and \texttt{3DSC}$_{\rm MP}$ is given in (b). Panels (a, b) and (c) were taken from Ref. \citenum{sommer20233dsc} (under the Creative Commons CC BY license) and Ref. \citenum{choudhary2022designing} (under the Creative Commons CC BY license), respectively.}\label{fig:ml10}
\end{figure*}

Some approaches were used to bring in atomic-level information. First, the chemical formula of a superconductors are looked up in the available databases like ICSD \cite{bergerhoff1983inorganic} and Materials Project \cite{Jain2013} for the most-reasonable structures \cite{isayev2015materials,stanev2018machine,zhang2022machine}. Fig. \ref{fig:ml10} (a) details the procedure \cite{sommer20233dsc} used to create \texttt{3DSC}$_{\rm ICSD}$ and \texttt{3DSC}$_{\rm MP}$, two superconductor datasets in which $T_{\rm c}$ from \texttt{NIMS SuperCon} is paired with the atomic structures from ICSD and Materials Project by matching the chemical formula. In this procedure, an ``artificial doping'' step was used to obtain complete matches in the chemical formula from ``nearly-complete'' matches. A summary of \texttt{3DSC}$_{\rm ICSD}$ and \texttt{3DSC}$_{\rm MP}$ is given in Fig. \ref{fig:ml10} (b).

\begin{table*}[ht]
  \caption{\footnotesize A summary of the currently available databases of superconductors that have been/can be used for machine-learning techniques.}\label{table:data}
\begin{footnotesize}
  \begin{tabular}{p{2.3cm} p{7.3cm} p{6.1cm} p{1.4cm}}
    \hline
  Name & Description  & URL (\texttt{https://+}) & References\\
    \hline
  \texttt{NIMS SuperCon} & $\simeq 31,000$ records of chemical formula \& $T_{\rm c}$ @ 0 GPa &supercon.nims.go.jp/index\_en.html& \\
  \texttt{MDR SuperCon} & Originated from \texttt{NIMS SuperCon}, $26,323$ records of chemical formula \& $T_{\rm c}$ @ 0 GPa &mdr.nims.go.jp/collections/5712mb227& \\
  \texttt{SuperCon} & Originated from \texttt{NIMS SuperCon}, $16,400$ records of chemical formula \& $T_{\rm c}$ @ 0 GPa & github.com/vstanev1/Supercon & ~\citenum{hamidieh2018data, stanev2018machine}\\
  \texttt{3DSC}$_{\rm ICSD}$ & $9,150$ records of atomic structure @ $0$ GPa (from ICSD) ``matched'' with experimental $T_{\rm c}$ (from \texttt{NIMS SuperCon}) via chemical formula. License needed for ICSD. & github.com/aimat-lab/3DSC& ~\citenum{sommer20233dsc} \\
  \texttt{3DSC}$_{\rm MP}$ & $5,759$ records of atomic structure @ $0$ GPa (from Materials Project) ``matched'' with experimental $T_{\rm c}$ (from \texttt{NIMS SuperCon}) via chemical formula& github.com/aimat-lab/3DSC& ~\citenum{sommer20233dsc} \\
  \texttt{Jarvis\_EPC} & $626$ records of atomic structure and $\lambda$ \& $\omega_{\rm log}$, computed @ $0$ GPa using DFPT & doi.org/10.6084/m9.figshare.21370572& ~\citenum{choudhary2022designing} \\
  N/A & $\simeq 7,000$ records of atomic structure and $\lambda$ \& $\omega_{\rm log}$, computed @ $0$ GPa using DFPT & & ~\citenum{cerqueira2024sampling} \\
  \texttt{CompSC} & $587$ atomic structures for which $584$ values of $\lambda$ \& $567$ values of $\omega_{\rm log}$ were computed @ up to 500 GPa. Data from literature, reoptimized using DFT, \& validated visually & github.com/huantd/matsdata & ~\citenum{tran2023machine} \\
  \texttt{SC-CoMIcs} & 1,000 annotated abstracts, developed \& tailored for extracting superconductivity-related information, e.g., using NLP& data.mendeley.com/datasets/xc9fjz2p3h/2 github.com/tti-coin/sc-comics& ~\citenum{yamaguchi2020sc, yamaguchi2022superconductivity} \\
  \texttt{SuperMat} & $142$ articles, $16,052$ entities, \& $1,398$ links, for NLP & github.com/lfoppiano/SuperMat& ~\citenum{foppiano2021supermat}\\
  \texttt{SuperCon$^2$} & $40,324$ records of superconductors, $T_{\rm c}$, applied pressure, measurement method & github.com/lfoppiano/supercon& ~\citenum{foppiano2023automatic}\\
    \hline
  \end{tabular}
\end{footnotesize}
\end{table*}

In the second approach, superconductivity-related parameters such as $\alpha^2F(\omega)$, $\lambda$, $\omega_{\rm log}$, and $T_{\rm c}$ are computed (see Sec. \ref{sec:comput}) for the atomic structures obtained from existing databases and/or predicted computationally \cite{shipley2021high, choudhary2022designing, cerqueira2024sampling}. \texttt{JARVIS-EPC} is a dataset generated \cite{choudhary2022designing} using such a computationally demanding approach with technical details shown in Fig. \ref{fig:ml10} (c). Starting from \texttt{JARVIS-DFT}, a database of 55,645 materials, screening steps involving some accessible data, e.g., Debye temperature and $N_{\rm F}$, were used \cite{choudhary2022designing}. Then, $\alpha^2F(\omega)$, $\lambda$, and $\omega_{\rm log}$ were computed for 1,058 materials, identifying 626 dynamically stable structures, 105 of them have McMillan $T_{\rm c} \geq 5$ K. More recently, a dataset of more than 7,000 records of atomic structures and their computed $\lambda$, $\omega_{\rm log}$, and $T_{\rm c}$ was obtained in a ML-driven computational discovery effort, whose workflow is shown in Fig. \ref{fig:ml4} (e) \cite{cerqueira2024sampling}. The two computational datasets are summarized in Table \ref{table:data}.

The third approach is inspired by the presence of thousands of computational reports for possible superconductivity at multiple ranges of pressure. Most of them start from the atomic structures predicted computationally, thus being highly expensive and trustworthy. The main challenge in this approach is how to collect and validate the literature data in a reliable and scalable manner. In an initial effort, a few hundred atomic structures and their $\lambda$ and $\omega_{\rm log}$, computed at pressures up to 500 GPa, were manually collected \cite{tran2023machine}. The curation involves constructing the reported atomic structures, uniformly optimizing them using DFT, and inspecting them visually. The resulted dataset \texttt{CompSC}, summarized in Table \ref{table:data}, contains $587$ atomic structures for which $584$ values of computed $\lambda$ and $567$ values of computed $\omega_{\rm log}$ are available \cite{tran2023machine}. This approach can create a reliable and highly diverse data, but laborious and obviously unscalable.

Using Natural Language Processing (NLP) tools to automatically extract superconductor-related data from scientific literature, a largely unexplored data reservoir, is more scalable and sustainable. This approach has emerged and recently gained some momentum \cite{foppiano2019proposal,yamaguchi2020sc, yamaguchi2022superconductivity,foppiano2021supermat,foppiano2023automatic}. \texttt{SC-CoMIcs} is a corpus of 1,000 annotated abstracts, created \cite{yamaguchi2020sc, yamaguchi2022superconductivity} for extracting superconductivity-related information using NLP-based tools like Named Entity Recognition (NER). Going beyond abstracts, \texttt{SuperMat} is an annotated corpus, supplying $142$ full-texts, which contain $16,052$ entities and $1,398$ links \cite{foppiano2021supermat}. Such efforts were further levitated to \texttt{Grobid-superconductors}, a module designed to automatically extract superconductor names and properties from text, and finally to \texttt{SuperCon$^2$}, a database containing $40,324$ records of superconductors chemical formulae, $T_{\rm c}$, applied pressure, and measurement method \cite{foppiano2023automatic}.

\subsection{Categorizing superconductors from data}\label{sec:categ}
\begin{figure}[t]
\centering
\includegraphics[width=0.950\linewidth]{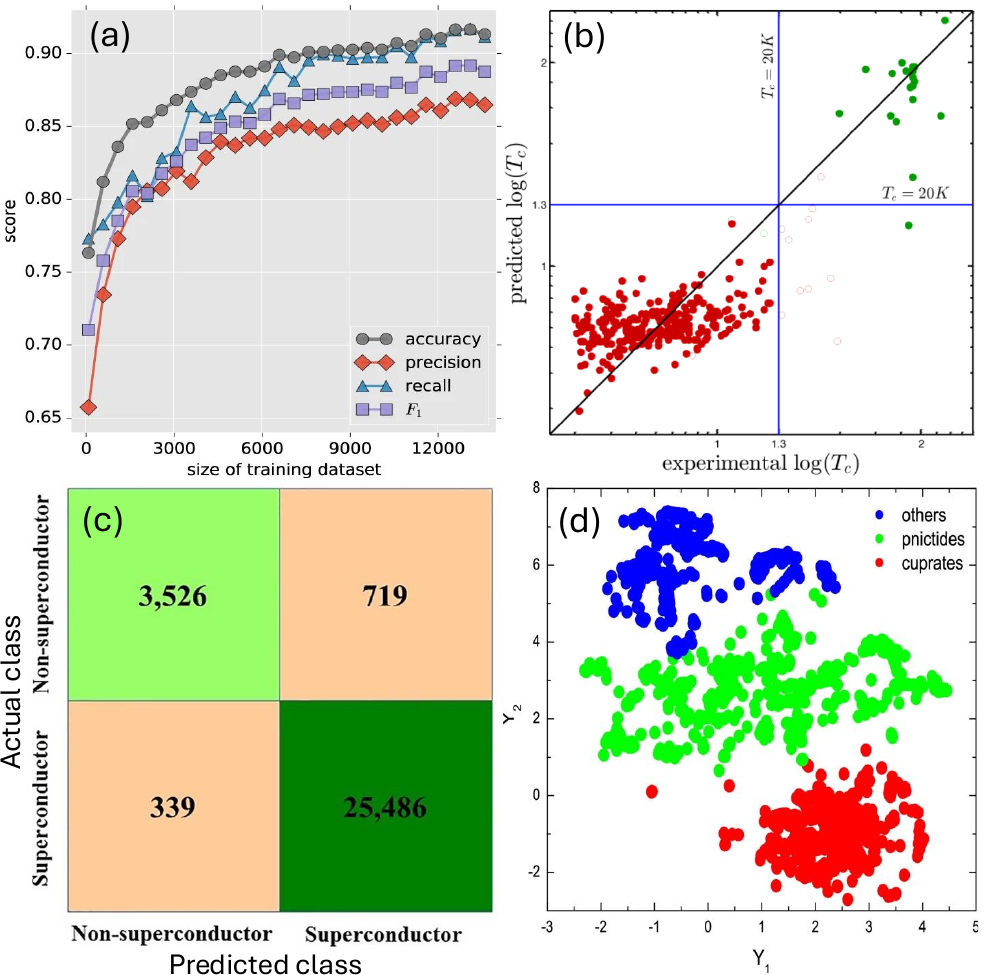}
\caption{(a) Four accuracy-related score of the superconductor/non-superconductor classification model, (b) predictions of a classification model (given in red and green labels) and a regression model (given on the $y$ axis) on a dataset of 464 structure-$T_{\rm c}$ records, (c) a confusion matrix for the classification model for superconductors that can reach an accuracy of 96.5\%, and (d) t-SNE plot of 4,500 randomly selected superconductors from the \texttt{SuperCon}, showing distinct clusters for different classes of superconductors. Panels (a), (b), (c), and (d) were reprinted with permission from Ref. \citenum{stanev2018machine} (under the Creative Commons CC BY license), Ref. \citenum{isayev2015materials} (Copyright 2015, American Chemical Society), Ref. \citenum{roter2020predicting} (Copyright 2020, Elsevier), and Ref. \citenum{roter2022clustering} (Copyright 2022, Elsevier), respectively.}\label{fig:classif}
\end{figure}

The developed databases could be useful for two categorical questions. First, whether or not a material is a superconductor, and second, if yes, what class, e.g., cuprates and iron-based, it belongs to. In an endeavor to address the first question, \texttt{SuperCon} (with 16,400 records) was augmented by 300 materials found \cite{hosono2015exploration} to be non-superconducting. For these materials, $T_{\rm c}$ was set to zero \cite{stanev2018machine}. Then, an adjustable parameter $T_{\rm sep}$ was introduced to separate the combined dataset into two groups. The ``below-$T_{\rm sep}$'' group includes the non-superconductors ($T_{\rm c} = 0$ K), $\simeq 4,000$ records without $T_{\rm c}$, and those with $T_{\rm c} < T_{\rm sep}$, while the ``above-$T_{\rm sep}$'' group hosts records with $T_{\rm c} \geq T_{\rm sep}$. By using the Magpie features \cite{ward2016general} for chemical formulae, setting $T_{\rm sep}=10$ K, and employing Random Forest algorithm \cite{breiman2001random}, the classification model developed, shown in Fig. \ref{fig:classif} (a), can reach an accuracy of $\simeq 92\%$ \cite{stanev2018machine}.

By including non-superconducting materials in the ``below-$T_{\rm sep}$'' group, the problem of superconductor recognition was just partially addressed \cite{stanev2018machine}. The reason is that $T_{\rm c}$ is not the only measurable characteristics of a superconductor but the only superconductivity-related property available in \texttt{NIMS SuperCon} dataset. Moreover, as $T_{\rm c}$ may be arbitrarily low \cite{kohn1965new}. some materials currently classified as non-superconducting may be discovered to exhibit superconductivity as technological advancements enable us to probe lower temperature regimes. An example of such a (very rare) finding is the discovery of superconductivity in elemental Li below $4\times 10^{-3}$ K \cite{tuoriniemi2007superconductivity}, despite a long-standing belief that Li would not exhibit superconductivity \cite{matthias1957chapter, buzea2004assembling}.

In fact, the first attempt to categorize superconductors based on $T_{\rm c}$ emerged slightly earlier, starting from $\simeq 700$ chemical formula-$T_{\rm c}$ records collected from literature, handbooks, and \texttt{NIMS SuperCon} \cite{isayev2015materials}. The chemical formulae were then matched with suitable atomic structures in {\sc aflowlib} \cite{AFLOWLIB}, creating a dataset of 464 structure-$T_{\rm c}$ records. The data were featurized by SiRMS, a fragment-based Simplex representation \cite{kuz2008hierarchical} before learned using the Random Forest algorithm. By setting the temperature separator $T_{\rm sep} = 20$ K, an accuracy of $\simeq 0.97$ was search by the obtained classification model, visualized in Fig. \ref{fig:classif} (b).

In an effort to distinguish superconductors from non-superconductors, \texttt{NIMS SuperCon} was cleaned and augmented by 3,000 non-superconducting materials, including insulators, semiconductors, and metals \cite{roter2020predicting}. Then, each chemical formula was represented by a row vector formed by the composition (contribution) of the constituent species. These vectors were aggregated in a chemical composition matrix that has 96 columns for 96 available species and $\simeq 30,000$ rows for the dataset size. Fig. \ref{fig:classif} (c) shows that the classification model trained using the k-Nearest Neighbors algorithm performs very well \cite{roter2020predicting} with an overall accuracy of $\simeq 96$\%. It is worth noting that the augmented 3,000 materials were assumed \cite{roter2020predicting} to be non-superconductors, skipping the aforementioned small possibility that some (metals) of them may be superconductors at very low temperatures.

The second question involves recognizing superconductors of different classes, i.e., those governed by different pairing mechanisms. Addressing this problem, \texttt{NIMS SuperCon} was cleaned and represented \cite{roter2022clustering} by the chemical composition matrix \cite{roter2020predicting}. Several clustering algorithms were tested before Self-Organizing Map \cite{kohonen2013essentials} was selected. Fig. \ref{fig:classif} (d) shows the $t$-distributed Stochastic Neighbor Embedding ($t$-SNE) \cite{van2008visualizing} of 4,500 superconductors randomly selected from \texttt{SuperCon}, in which iron-based, cuprates, and those in other classes are distinguished \cite{roter2022clustering}. It seems that data, when curated and learned properly, could be useful to recognize a superconductor and the governing mechanism, if applicable.

\subsection{Predicting superconducting-related properties}
ML efforts in this class aim at accelerating the predictions of superconducting-related properties such as $T_{\rm c}$, traditionally obtained by expensive computations (Sec. \ref{sec:comput}) and/or physical measurements (Sec. \ref{sec:expt}). The critical role of \texttt{NIMS SuperCon} and its descendants, evidenced in Sec. \ref{sec:categ}, is also visible here. In fact, most of the works aiming at predicting $T_{\rm c}$ from the chemical formula rely on \texttt{NIMS SuperCon}. Efforts to introduce atomic-level information in the developent of ML models for $\lambda$ and $\omega_{\rm log}$ emerged recently. This Section is devoted to not only the ML efforts in the two subcategories but also the ML-driven searches for superconductors. An in-depth discussion on the remaining challenges and opportunities of ML efforts in this class is given in Sec. \ref{sec:deeplearning}.

\subsubsection{Predictions from chemical formula}
\begin{figure}[t]
\centering
\includegraphics[width=0.950\linewidth]{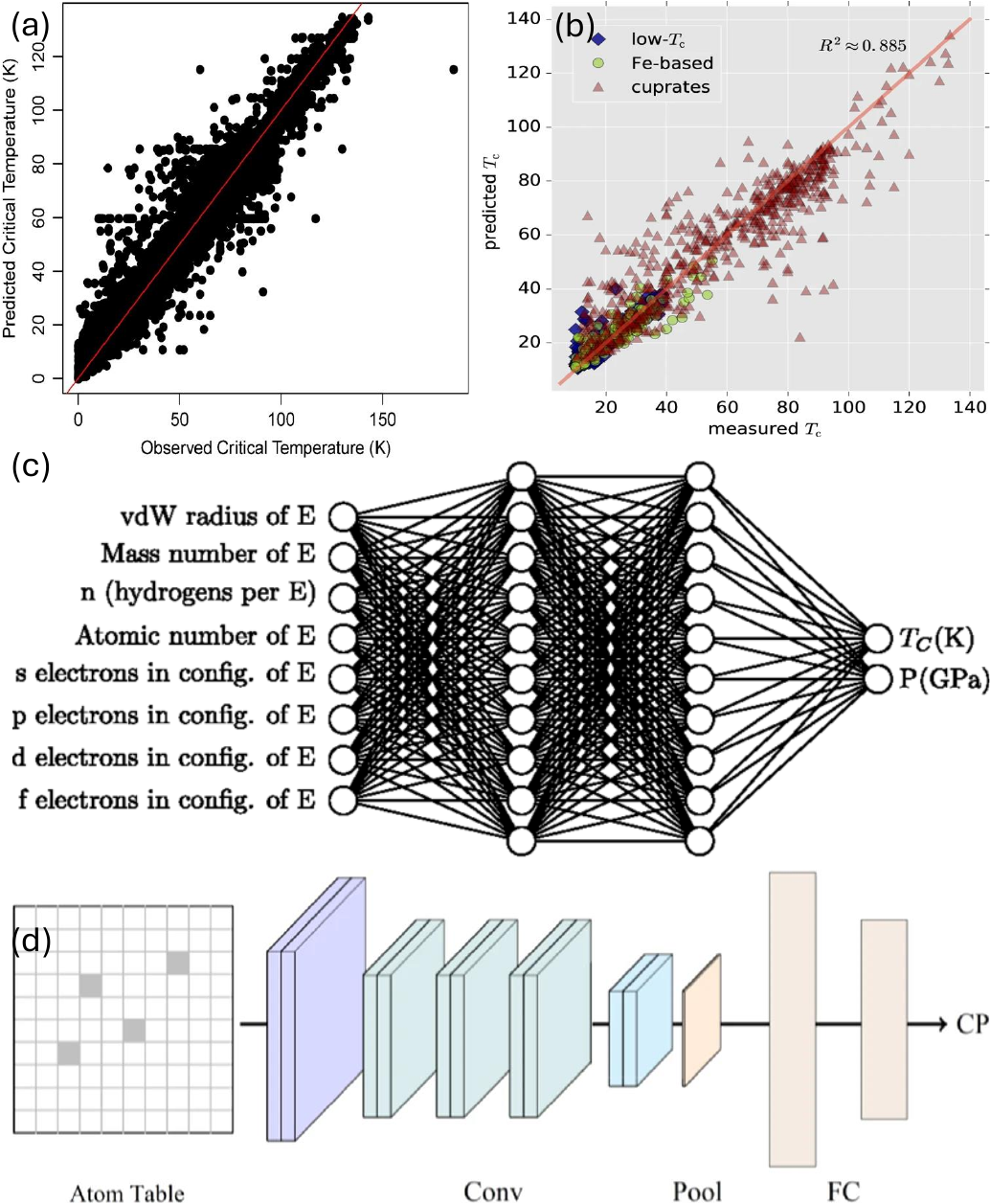}
\caption{(a, b) Two ML models for predicting $T_{\rm c}$ from chemical formula, (c) a Fully-Connected Neural Network created to predict $T_{\rm c}$ and $P$ from chemical formula, and (d) an atom table Convolutional Neural Network used to train ML predictive model for $T_{\rm c}$ on the datasets whose chemical formulae are represented as $10\times 10$ images. Panels (a), (b), (c), and (d) were reprinted with permission from Ref. \citenum{hamidieh2018data} (Copyright 2018, Elsevier), Ref. \citenum{stanev2018machine} (under the Creative Commons CC BY license), Ref. \citenum{hutcheon2020predicting} (Copyright 2020, American Physical Society), and Ref. \citenum{zeng2019atom} (under the Creative Commons CC BY license), respectively.}\label{fig:ml2}
\end{figure}

In an early ML work starting from \texttt{NIMS SuperCon}, a dataset of 21,263 records was curated, containing multiple classes of superconductors \cite{hamidieh2018data}. For each material, the chemical formula was featurized by some simple functions, e.g., mean, weighted mean, and entropy, etc., of the basic properties of the constituent species, e.g., atomic mass and electron affinity, etc. Some algorithms were tested and XGBoost \cite{chen2016xgboost} was selected. The developed ML model for $T_{\rm c}$ with an averaged out-of-sample error of $\simeq 9.5$ K is visualized in Fig. \ref{fig:ml2} (a) \cite{hamidieh2018data}.

Following the classification between low-$T_{\rm c}$ and high-$T_{\rm c}$ superconductors, $T_{\rm c}$ predictive models were developed for the high-$T_{\rm c}$ group \cite{stanev2018machine}. The materials data were represented by Magpie (composition) features \cite{ward2016general} and learned by the Random Forest (regression) algorithm. One of the models developed is visualized in Fig. \ref{fig:ml2} (b), showing the performance on low-$T_{\rm c}$, Fe-based, and cuprate superconductors with a coefficient of determination $R^2 \simeq 0.88$ \cite{stanev2018machine}. Screening ICSD database by a combination of a classification and a regression model, $\simeq 2,000$ materials with predicted $T_{\rm c} > 20$ K at 0 GPa were identified \cite{stanev2018machine}. While most of them contain copper and oxygen, i.e., they may be related to cuprates, a subset of 35 materials without obvious connection to known high-$T_{\rm c}$ families was compiled and reported \cite{stanev2018machine}.

Roughly 2,000 records of AlB$_2$-, Chevrel-, Cr$_3$Si-, spinel-, NaCl-, and skutterudite-type superconductors were extracted from \texttt{NIMS SuperCon} and augmented with elemental superconductors \cite{matsumoto2019acceleration}. This dataset was believed to include phonon-mediated superconductors, for which MgB$_2$ \cite{nagamatsu2001superconductivity} is the highest-$T_{\rm c}$ material -- the averaged $T_{\rm c}$ of MgB$_2$ in this dataset is 38.6 K. Featurizing the data by some simple functions of the fundamental attributes of existing species, the trained Random Forest model could reach an $R^2 = 0.98$ on the training data (80\% of the dataset) and $R^2 = 0.92$ on the test data (the remaining 20\% of the dataset). This model was used to create predicted $T_{\rm c}$ map for three family of ternary materials, namely Mg-B-Ti, Fe-Te-Se, and Ca-B-C. In the Mg-B-Ti map, the region containing MgB$_2$ was predicted to be highest in $T_{\rm c}$ while in the Fe-Te-Se map, the high $T_{\rm c}$ was predicted along the line connecting FeTe and FeSe \cite{matsumoto2019acceleration}. For the Ca-B-C system, materials with formula close to CaB$_6$ and B$_{13}$C$_2$ were also predicted \cite{matsumoto2019acceleration} to be high in $T_{\rm c}$.

Focusing on the recent discoveries, a few hundreds of binary hydride superconductors EH$_n$ were collected from literature \cite{hutcheon2020predicting}. For each material, its chemical formula, $T_{\rm c}$, and the pressure at which the superconductivity was predicted, are available. A set of features, including the hydrogen content $n$ and some fundamental attributes of the species E such as the atomic number, the van der Waals radius, and the electron configurations of E was used to describe EH$_n$. The featurized data was then fed into a Fully-Connected Neural Network, visualized in Fig. \ref{fig:ml2} (c), whose output layer has two nodes, one for $T_{\rm c}$ and the other for $P$. This network is an example of a multi-task learning architecture, which can be trained multiple datasets to exploit the hidden correlations among them and leverage the performance \cite{kuenneth2021polymer,kuenneth2022bioplastic, Tuoc:PrNN}. The $T_{\rm c}$ predictivity of the model reaches $R^2 = 0.88$ with root mean square error (RMSE) $\simeq 33.7$ K for $T_{\rm c}$ predictions.

Nevertheless, the main objective of the developed Neural Network model is to screen over the Periodic Table for the species E that minimize the ``distance'' from the predicted $T_{\rm c}$ and $P$ to the ambient conditions, i.e., $P=0$ GPa and $293$ K \cite{hutcheon2020predicting}. The analysis suggests that alkali and alkaline-earth metal hydrides could be the best candidates for superconductivity near ambient conditions. Next, the AIRSS method \cite{AIRSS} was used, identifying dozens of atomic structures of alkali and alkaline earth metal hydrides with respectable computed $T_{\rm c}$. Specifically for the predicted $C2/m$, $Cmcm$, and $Immm$ phases of RbH$_{12}$, the computed $T_{\rm c}$ could be as high as 126 K at 100 GPa and below \cite{hutcheon2020predicting}.

Materials informatics endeavors relying on \texttt{NIMS SuperCon} were extended into the deep-learning territory, where high-level features may be learned directly from raw data. In the atom table Convolutional Neural Network (ATCNN) \cite{zeng2019atom}, each chemical formula was represented by an image of $10\times 10$ pixels, visualized at the left end of Fig. \ref{fig:ml2} (d). Each pixel corresponds to a species, and its value is the composition (contribution) of this species in the chemical formula. As there are 86 species that appear in the dataset, thus $10\times 10 = 100$ pixels is sufficient for the representation. Then, the ``atom tables'' are accepted by an architecture, sketched in Fig. \ref{fig:ml2} (d), which consists several convolutional layers to process the images \cite{zeng2019atom}, Two models for $T_{\rm c}$, i.e., ATCNN-I and ATCNN-II, were developed, one trained on a cleaned version of \texttt{NIMS SuperCon} containing 13,598 records and the other trained on the same dataset after being augmented by 9,399 energetically stable insulators, the obvious non-superconductors. Both models show good performance with a mean absolute error (MAE) of $\simeq 4.2$ K, RMSE of $\simeq 8.2$ K, and $R^2\simeq 0.97$. Compared to the measured $T_{\rm c}$ of some well-known superconductors such as Hg, MgB$_2$, and YBa$_2$Cu$_3$O$_7$, the predictions of these two models are accurate \cite{zeng2019atom}.

The idea of using a Convolutional Neural Network (CNN) architecture in superconductor discovery has evolved from recognizing ``atom tables'' \cite{zeng2019atom} to ``reading periodic tables'' \cite{konno2021deep}. In the latter, the species composition of each superconductor in \texttt{NIMS SuperCond} is ``written'' directly to the Periodic Table. Then, the table was separated into four ``channels'' for recording those with $s$, $p$, $d$, and $f$ valence electrons. A CNN was trained on 95\% of \texttt{NIMS SuperCond} and tested on the remaining 5\% of the data, yielding $R^2 = 0.92$ in $T_{\rm c}$ predictions \cite{konno2021deep}. This model predicts about $17,000$ materials (out of $\sim 48,000$ records) in Crystallography Open Database (COD) \cite{Grazulis2012}, to have $T_{\rm c}> 10$K. The obtained result is unreasonable, perhaps because the training data have almost no non-superconductors. After augmenting it by a synthetic dataset of non-superconductors (assumed $T_{\rm c}=0$ K), the final (new) model becomes more reliable. Testing on 400 materials (including 330 non-superconductors) reported in Ref. \citenum{hosono2015exploration}, the model reaches a precision of $62$\%, an accuracy of 76\%, a recall of 67 \%, and an f1 score of 63\% in predicting materials with $T_{\rm c}> 0$ K. Using it for COD, 70 materials, including CaBi$_2$ and Hf$_{0.5}$Nb$_{0.2}$V$_2$Zr$_{0.3}$ (both are not in \texttt{NIMS SuperCond}) were predicted to have $T_{\rm c}> 10$ K. One of them, CaBi$_2$ is indeed a superconductor \cite{winiarski2016superconductivity}.

We close this Section by noting the connection from the ``atom table'' and the ``periodic table'' in these CNN approaches to the chemical composition row vector \cite{roter2020predicting} discussed in Sec. \ref{sec:categ}. In particular, the former is the latter rearranged in a two-dimensional image so that it can be used in a CNN architecture. The main information incoded in these images is the composition of the constituent species, the highest level of information that can be extrated from a chemical formula.

\subsubsection{Predictions from atomic structure}
In fact, attempts to bring atomic-level details to ML models for $T_{\rm c}$ emerged quite early \cite{isayev2015materials}. By matching $\simeq 700$ formulae collected for superconductors with AFLOWLIB and excluding those with $T_{\rm c} <2$ K, a dataset of 295 atomic structure-$T_{\rm c}$ records was obtained. The atomic structures were represented using SiRMS \cite{kuz2008hierarchical} before being mapped onto $T_{\rm c}$ by Random Forest and Partial Least Squares \cite{wold2001pls} algorithms. The obtained models for $T_{\rm c}$ could reach $R^2 \simeq 0.66$, one of them is shown in Fig. \ref{fig:classif} (b). As SiRMS is a fragment-based representation \cite{kuz2008hierarchical}, an analysis was performed, compiling a catalog of fragments that may likely present in materials with low and high values of $T_{\rm c}$ \cite{isayev2015materials}.

\begin{figure*}[t]
\centering
\includegraphics[width=0.7\linewidth]{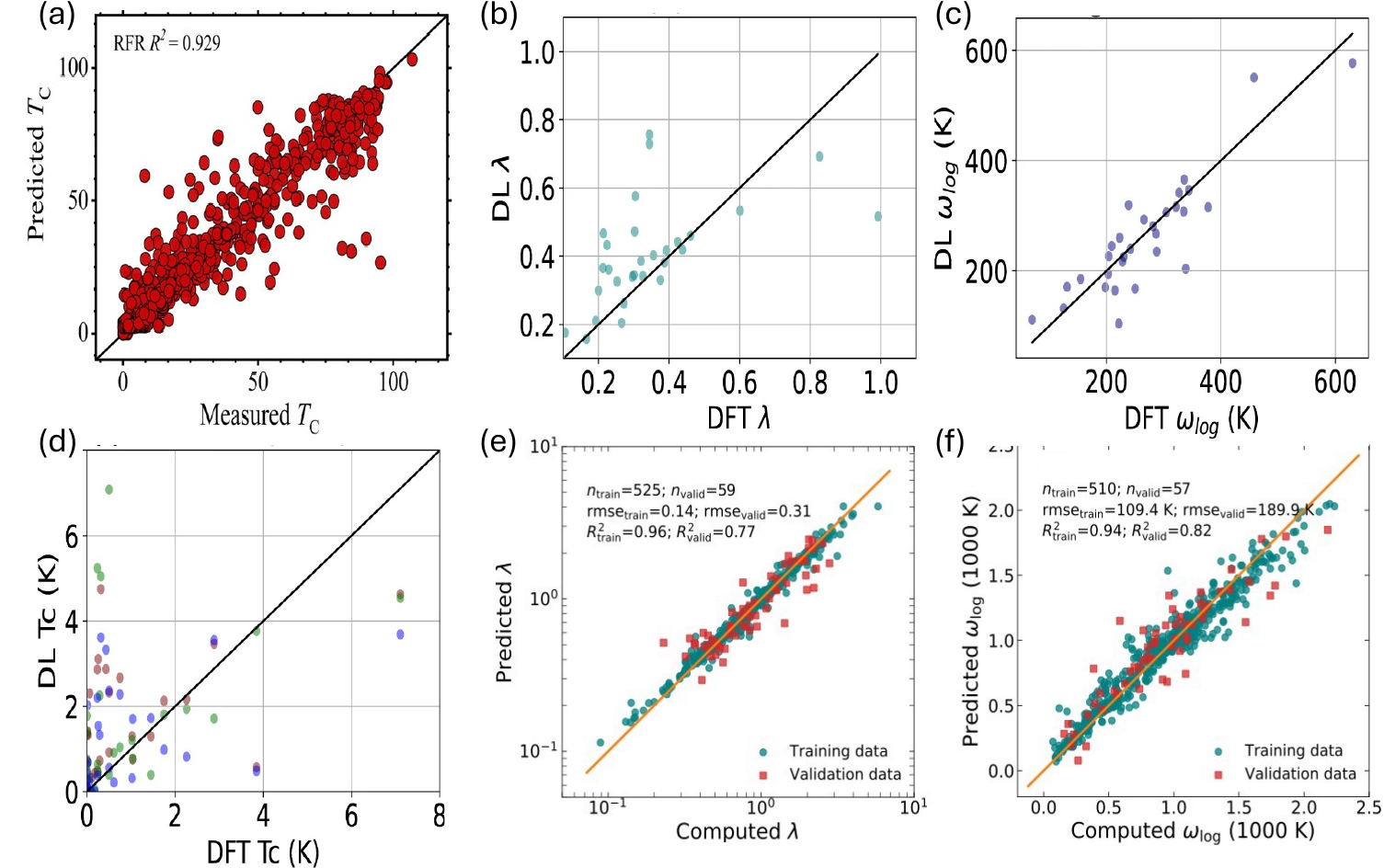}
\caption{Nine ML predictive models starting from atomic structures to predict (a, d) $T_{\rm c}$, (b, e) $\lambda$, and (c, f) $\omega_{\rm log}$. Panels (a), (b, c, d), and (e, f) were reprinted with permission from Ref. \citenum{zhang2022machine} (Copyright 2019, American Chemical Society), Ref. \citenum{choudhary2022designing} (under the Creative Commons CC BY license), and \citenum{tran2023machine} (Copyright 2023, American Physical Society), respectively.}\label{fig:ml3}
\end{figure*}

An atomic structure matching procedure was also used in Ref. \citenum{zhang2022machine}. By looking at ICSD database for which atomic structures are available, $\simeq 1,700$ superconductors were obtained. Next, by identifying suitable (close-matched) structures and ``doping'' them to recover the chemical formulae in \texttt{NIMS SuperCon}, a dataset of 5,713 atomic structure-$T_{\rm c}$ records was finally obtained \cite{zhang2022machine}. Then, the atomic structures were featurized using the ``Smooth Overlap of Atomic Position'' (SOAP) scheme \cite{de2016comparing}. Finally, three models were trained used Random Forest, XGBoost, and Support Vector Regression \cite{smola2004tutorial}. One of them, the Random Forest regression model with $R^2\simeq 0.92$, is visualized in Fig. \ref{fig:ml3} (a) \cite{zhang2022machine}. These models were used to screen over ICSD database, compiling 10 most promising superconductors whose averaged $T_{\rm c}$ (predicted by 3 models) is at least $\simeq 20$ K at 0 GPa \cite{zhang2022machine}. Among them, Ba$_4$Ca$_4$Cu$_6$O$_{19}$Tl$_3$ was predicted to have $T_{\rm c}\simeq 103$ K \cite{zhang2022machine}.

Starting from \texttt{JARVIS-EPC}, two learning methods were used to develop predictive models for $\lambda$, $\omega_{\rm log}$, and $T_{\rm c}$ \cite{choudhary2022designing}. The first relies on some force-field-inspired features and the Gradient Boosting Decision tree algorithm \cite{friedman2001greedy}. The second, referred to as ``Atomistic Line Graph Neural Network'' (ALIGNN), is a deep-learning architecture, in which a graph convolutional layer was designed to describe two- and three-body interactions among the atoms of an atomic structure \cite{choudhary2021atomistic}. Figs. \ref{fig:ml3} (b), (c), and (d) visualize three ALIGNN models \cite{choudhary2022designing} trained on \texttt{JARVIS-EPC} to predict $\lambda$, $\omega_{\rm log}$, and $T_{\rm c}$, respectively.

Two ML models trained \cite{tran2023machine} on \texttt{CompSC} to predict $\lambda$ and $\omega_{\rm log}$ are shown in Figs. \ref{fig:ml3} (e) and (f). In this work, the atomic structures predicted at pressures up to 500 GPa and reported in literature were collected, uniformly reoptimized, visually validated, represented using {\sc matminer} \cite{WARD201860}, and learned using Gaussian Process Regression (GPR) algorithm \cite{GPR95, GPRBook}. The training data are highly diverse, containing numerous (``unusual'') atomic details realized at different $P$ and computationally linked to the values of $\lambda$ and $\omega_{\log}$ that lead to high values of $T_{\rm c}$. Therefore, the models are expected to be capable of recognizing at any $P$, including $0$ GPa, the atomic structures that resemble the ``unusual'' atomic-level details they were exposed to \cite{tran2023machine}. Using these models to screen the hydrides from Materials Project database, an $Fm\overline{3}m$ structure of CrH and another $Fm\overline{3}m$ structure of CrH$_2$ were identified with computed $T_{\rm c} = 15.7$ K and $10.7$ K, respectively, at $0$ GPa \cite{tran2023machine}.

\subsubsection{ML-driven search for high-$T_{\rm c}$ superconductors}\label{sec:mldisc}
\begin{figure*}[t]
\centering
\includegraphics[width=0.650\linewidth]{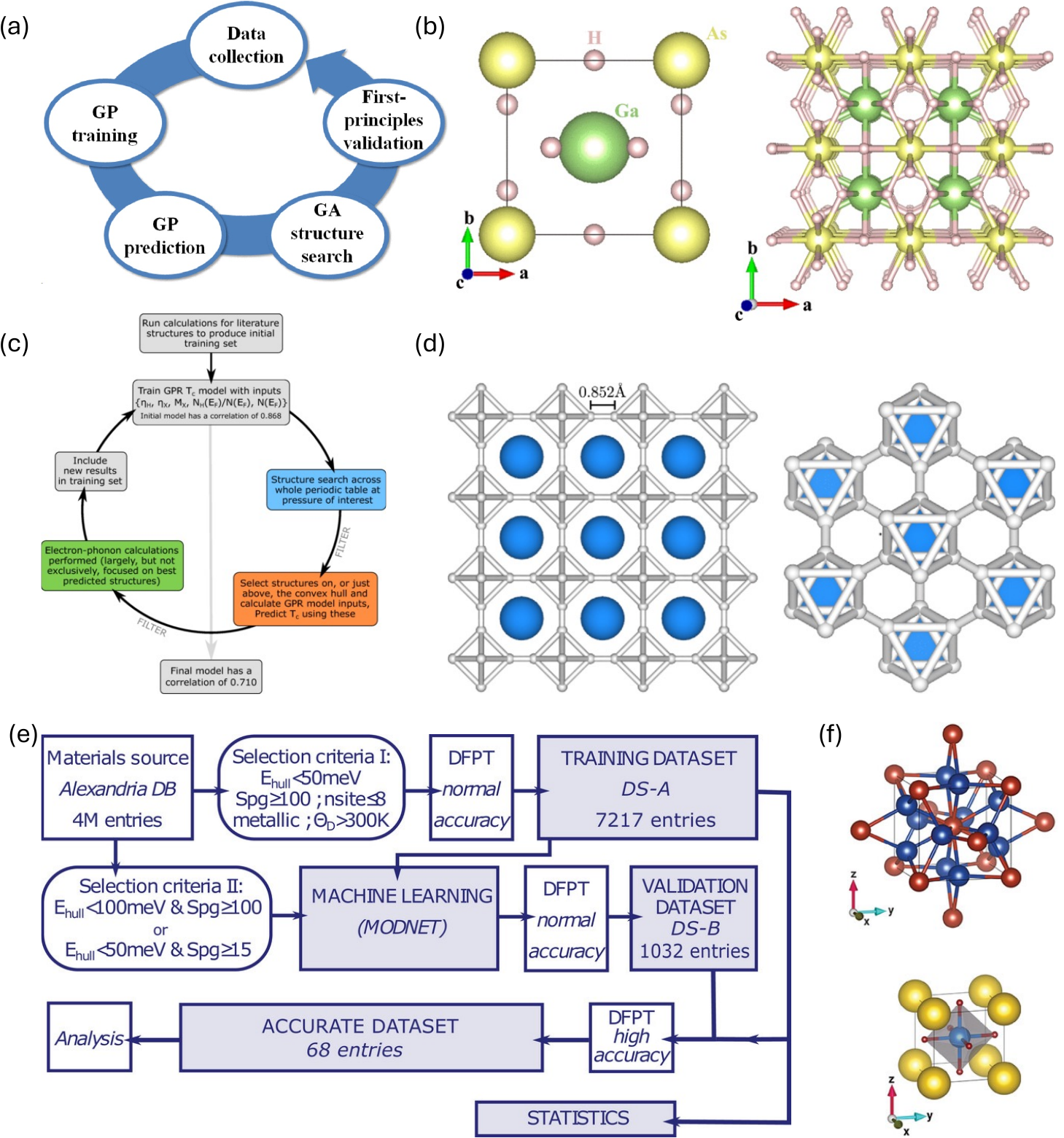}
\caption{(c) A discovery loop using a ML predictive model to search for high-$T_{\rm c}$ superconductors, (d) the atomic structure of discovered GaAsH$_6$, (e) the atomic structure of discovered NaH$_6$, and (f) the \'{E}liashberg spectral function $\alpha^2F(\omega)$ of NaH$_6$ and some related materials. Panels (a, b), (c, d), and (e, f) were taken from Ref. \citenum{ishikawa2019materials} (under the Creative Commons Attribution 4.0 International license), Ref. \citenum{shipley2021high} (Copyright 2021, American Physical Society), and Ref. \citenum{cerqueira2024sampling} (Copyright 2024, John Wiley and Sons), respectively.}\label{fig:ml4}
\end{figure*}

An efficient materials discovery strategy, even powered by ML models, should be target-driven, extending beyond a brute-force screening, as discussed in Sec. \ref{sec:invdesign}. One such workflow was developed \cite{ishikawa2019materials}, utilizing an evolutionary algorithm and a GPR model to discover hydrogen-containing superconductors. In this ML-driven strategy, visualized in Fig. \ref{fig:ml4} (a), mating and mutation operate directly on the atomic structures while space group, hydrogen concentration, atomic mass, pressure, and $\mu^*$ were used as descriptors to train the GPR model \cite{ishikawa2019materials}. Having new candidates, new computed data come, the ML model is retrained, and the workflow cycles. Some hydride superconductors were discovered, including a $C2/m$ structure of KScH$_{12}$ with computed $T_{\rm c} = 122$ K at 300 GPa and a $Pm\overline{3}$ structure of GaAsH$_6$, shown in Fig. \ref{fig:ml4} (b), with computed $T_{\rm c} = 98$ K at 180 GPa.

Another ML-powered discovery strategy for hydride superconductors was recently developed \cite{shipley2021high} and visualized in Fig. \ref{fig:ml4} (c). In this procedure, the GPR model was trained on some formula-based features like the Gaspari-Gyorffy electron-phonon coupling estimates of available species, the atomic mass of non-hydrogen species, and $N_{\rm F}$; all of them were normalized appropriately. In this strategy, the atomic structures of the formula candidates were searched using AIRSS \cite{AIRSS}, and as the workflow cycles, 27 new superconductors were discovered at pressures up to 500 GPa. Among them, NaH$_6$, shown in Fig. \ref{fig:ml4} (d), was predicted to have $T_{\rm c}$ in the range from 228 K to 279 K at 100 GPa \cite{shipley2021high}.

Fig. \ref{fig:ml4} (e) visualizes the workflow used to create the dataset of $\simeq 7,000$ records discussed in Sec. \ref{sec:data} and summarized in Table \ref{table:data} \cite{cerqueira2024sampling}. In this workflow, a deep-learning model trained on a ``Material Optimal Descriptor Network'', or MODNet \cite{de2021materials}, was used to select candidates from Alexandria database \cite{AlexandriaDB} for $\lambda$, $\omega_{\rm log}$, and $T_{\rm c}$ computations. MODNet accepts features computed using {\sc matminer} \cite{WARD201860} for the atomic structures before passing them through a series of successive blocks for feature selection, encoding, decoding, and splitting to learn multiple properties concurrently. Among $\simeq 7,000$ data records produced using this workflow, 541 materials with $T_{\rm c} > 10$ K (at $0$ GPa) were identified. Two of them, a $Pm\overline{3}n$ structure of Ti$_3$Te (computed $T_{\rm c} = 16.3$ K) and a $Pm\overline{3}m$ structure of KCdH$_3$ (computed $T_{\rm c} = 12.3$ K) are visualized in Fig. \ref{fig:ml4} (f) \cite{cerqueira2024sampling}.

\subsection{Accelerated structure prediction with ML potentials}\label{sec:MLpotential}
In the computational discovery workflow discussed in Sec. \ref{sec:invdesign}, the structure prediction step is computationally very demanding. For each chemical formula, hundreds of thousands of energy evaluations are typically needed to search for its stable atomic structures, and they must be performed at a first-principles level. The development of ML potentials \cite{Behler:NNPot}, mostly in the context of molecular dynamics (MD) simulations, offer an approach to accelerate this step. A ML potential is a ML model, generally based on a Neural Network, accepting an atomic structure and returning its potential energy faster than a normal DFT calculation by multiple orders of magnitude. {\it State-of-the-art} ML potentials such as Gaussian Approximation Potential (GAP) \cite{Bartok:GAP1, bartok2017machine}, Spectral Neighbor Analysis Potential (SNAP) \cite{Thompson2015316, Chen:MoFF}, Moment Tensor Potentials (MTP) \cite{shapeev2016moment, gubaev2019accelerating}, and Deep Potential \cite{zhang2018deep}, trained on millions of atomic environments, are expected to be comparable with DFT in accuracy, i.e., a few meV/atom. However, like any ML models, out-of-domain energy predictions (on unseen atomic environments) should be treated with care. In superconductor discovery, ML potentials started leaving discernible trails \cite{gu2019superconducting, yang2021hard, dolui2024feasible, lucrezi2023quantum, ferreira2023search}.

A reliable atomic structure predicted computationally should be thermodynamically, dynamically, and kinetically stable. In the context of superconductor discovery, the dynamical stability of a structure, i.e., whether it is a local minimum, not a saddle point of the PES, is accessible during the calculations of the spectral function $\alpha^2F(\omega)$ using DFPT. Examining the kinetic stability of a structure, i.e., if the kinetic (energy) barrier protecting this local minimum is high enough so its lifetime could be reasonable, is much harder than inspecting its thermodynamic stability. The reason is that the former is non-local in nature, especially when the configuration space is extremely high in dimensionality, which is roughly 3 times of the system size, i.e., in orders of $10^2$.

In most of the cases, ML potentials were used to accelerate the energy evaluations during the search, i.e., their role is on the side of thermodynamic stability. Some results of these efforts are the discoveries at $P=0$ GPa of C$_4$K (in $P4/mmm$ symmetry with predicted $T_{\rm c} = 30.4$ K) \cite{gu2019superconducting}, $c$-B$_{\rm 24}$ (in $Pm\overline{3}$ symmetry with predicted $T_{\rm c} = 13.8$ K) \cite{yang2021hard}, and Mg$_2$IrH$_6$ (in $Fm\overline{3}m$ symmetry with predicted $T_{\rm c} = 160$ K) \cite{dolui2024feasible}. At a slightly higher pressure ($P=20$ GPa), a series of 14 La-N-H trinary materials was predicted with support from the universal Neural Network potential developed by {\sc matlantis} \cite{matlantis}. Examining their possible superconductivity by DFPT calculations \cite{ishikawa2024evolutionary}, the predicted $T_{\rm c}$ of this series spans from $0.49$ K to $14.41$ K. In another interesting effort \cite{ferreira2023search}, ephemeral data-derived (ML) potentials \cite{pickard2022ephemeral} were used in a large-scale structure prediction campaign, ending up disproving an early claim of ambient-condition superconductivity in the Lu-N-H systems. Compared with positive conclusions, such a negative conclusion requires {\it much bigger} computational effort because all the attainable possibilities, e.g., chemical formulae and system sizes, should be considered. ML potentials are particularly useful for this purpose.

ML potentials have also been used to accelerate the examination of the kinetic stability, which connects with kinetic barriers. A typical approach is to perform long first-principles MD simulations at room temperatures and above to examine the stability of the long-range order against elevated thermal energies \cite{gu2019superconducting}. Accessing directly to the kinetic barriers, e.g., using stochastic self-consistent harmonic approximation, requires sufficiently large supercells and numerous randomly displaced structures \cite{lucrezi2023quantum}. With help from MTP, such a computationally expensive analysis was completed, suggesting the stability of the $Fm\overline{3}m$ phase of BaSiH$_8$, whose predicted $T_{\rm c}$ is about 90 K \cite{lucrezi2023quantum}.

\subsection{Refining empirical formulae for $T_{\rm c}$}\label{sec:ml3}
The empirical McMillan formula (\ref{ADM0}) of $T_{\rm c}$ was developed \cite{McMillanTc} and refined \cite{dynes1972mcmillan, AllenTc} by ``manually learning'' small datasets generated from the \'{E}liashberg equations. This formula, and some other variants, are particularly useful in the search for new superconductors. Recent ideas emerged \cite{xie2019functional, xie2022machine}, suggesting that these formulae may be refined further using advanced symbolic ML techniques and bigger datasets of the solutions of the \'{E}liashberg equations, which become available thanks to new generations of computational infrastructures.

One such symbolic ML techniques is ``Sure Independence Screening and Sparsifying Operator'' (SISSO) \cite{ouyang2018sisso}. By defining some physically meaningful operators and functions of the primary variables, which are $\lambda$, $\omega_{\rm log}$, and $\mu^*$, millions of features (expressions) were generated. Training a linear regression model using these features with $L_0$ regularization, some expressions for $T_{\rm c}$ were obtained, among them the simplest version is \cite{xie2019functional}
\begin{equation}
T_{\rm c} = 0.0953\frac{\lambda^4\omega_{\rm log}}{\lambda^3 + \sqrt{\mu^*}}.
\end{equation}
In a subsequent effort \cite{xie2022machine}, a modified version of the McMillan formula (\ref{ADM0}) was assumed as
\begin{equation}
T_{\rm c} = \frac{f_\omega f_\mu \omega_{\log}}{1.2}\exp\left[-\frac{1.04(1+\lambda)}{\lambda-\mu^*(1+0.62\lambda)}\right]
\end{equation}
in which two prefactors $f_\omega$ and $f_\mu$ are learned from data to be
\begin{equation}
f_\omega = 1.92\frac{\lambda + \omega_{\rm log}/\overline{\omega_2}-\sqrt[3]{\mu^*}}{\sqrt{\lambda}\exp\left(\omega_{\rm log}/\overline{\omega_2}\right)}-0.08
\end{equation}
and
\begin{equation}
f_\mu = \frac{6.86\exp(-\lambda/\mu^*)}{1/\lambda-\mu^*-\omega_{\rm log}/\overline{\omega_2}}+1
\end{equation}
where \cite{AllenTc}
\begin{equation}
\overline{\omega_2} = \left[\frac{2}{\lambda}\int_0^\infty d\omega\omega\alpha^2F(\omega)\right]^{1/2}.
\end{equation}

Likewise, data obtained from the ML-powered discovery strategy developed \cite{shipley2021high} for hydride superconductors offer an opportunity to test and improve the McMillan formula (\ref{ADM0}). Assuming
\begin{equation}
T_{\rm c} = \frac{\omega_{\log}}{1.2}\exp\left[-\frac{1.04(1+\lambda)}{\lambda-\mu^*(1+0.62\lambda)}\right](a+b\lambda),
\end{equation}
data fitting yields $a = 1.0061$ and $b = 0.0663$ \cite{shipley2021high}, slightly modifying the McMillan formula (\ref{ADM0}).

\section{Challenges, opportunities, and critical next steps}\label{sec:challenge}
\subsection{Theoretical and computational methods}\label{sec:challenge_theory}
Despite thousands of computational discoveries reported in the last two decades, only about $20-30$ materials were synthesized and reported. This small fraction (about 1\% or below) may suggest that we simply do not have a ``silver bullet'' for such a tough problem like superconductivity understanding, prediction, and discovery. Looking forward, this situation may change, but some major challenges must be resolved. Apparently, phonon-mediated pairing is one of many mechanisms proposed for superconductivity, and an exact and/or well-controlled theory for this mechanism is not readily available. It is known that Migdal-\'{E}liashberg theory may become inaccurate \cite{esterlis2018breakdown, alexandrov2001breakdown, hague2008breakdown, benedetti1998holstein, yuzbashyan2022breakdown, dynes1986breakdown}, and one reason could be the fact that it accounts only for the first order of vertex corrections \cite{chubukov2020eliashberg}. Challenges in pushing the theoretical front in superconductivity understanding are obviously enormous.

In the computational front, two major steps of the workflow discussed in Sec. \ref{sec:invdesign}, namely the atomic structure prediction and the computations of $\alpha^2F(\omega)$, $\lambda$, and $\omega_{\rm log}$, are highly non-trivial. The former faces truly infinite, extremely high-dimensional configuration spaces. For the latter, implications of the possible breakdown of Migdal-\'{E}liashberg theory have not been well understood while the desirable accuracy and convergence are hard to control and obtain \cite{shipley2021high, choudhary2022designing, tran2023machine, cerqueira2024sampling}. Technically, even a numerical error that is comparable with the best attainable level of accuracy, i.e., $\sim 10^{-4} - 10^{-3}$ eV/\AA~ in the atomic force calculations, could be translated into a sizeable change in the computed phonon frequencies, and thus, in the superconductivity-related properties.

\begin{figure}[t]
\centering
\includegraphics[width=0.95\linewidth]{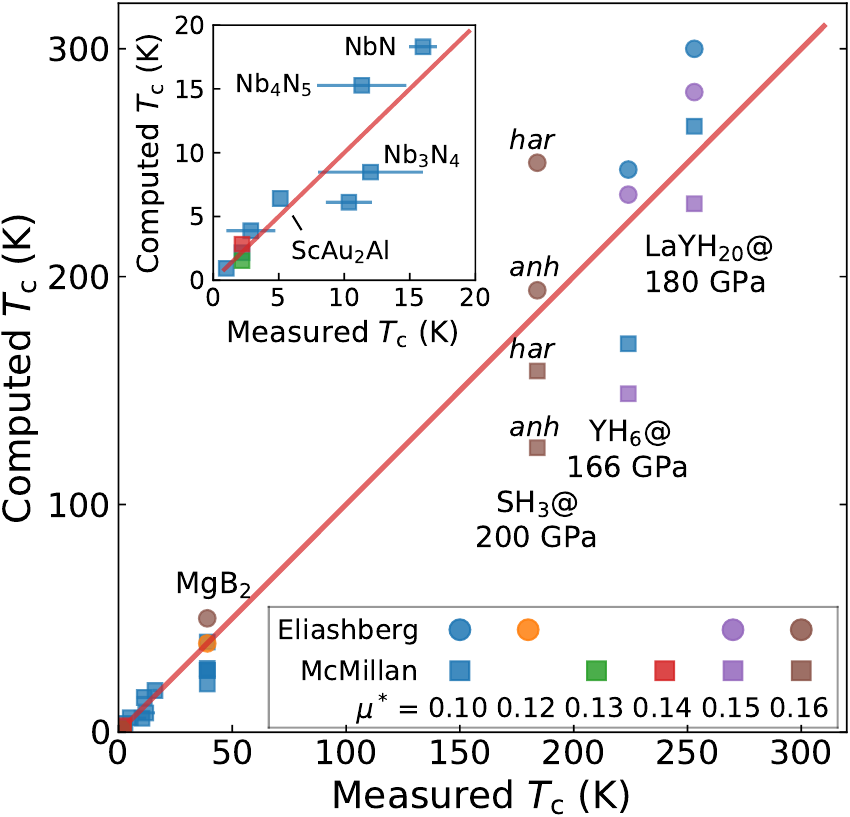}
\caption{The critical temperature $T_{\rm c}$, computed for some notable superconductors using McMillan formula and solving the \'{E}liashberg equations, given in comparison with measured $T_{\rm c}$. Value of $\mu^*$ and whether or not anharminicity, labeled by \texttt{anh} and \texttt{har}, is used, are shown. The inset is used for the small-$T_{\rm c}$ regime. The $x$-axis errror bars were obtained from different measured values of $T_{\rm c}$ while different computed values of $T_{\rm c}$ are shown separately for further discussions.}\label{fig:comp_valid}
\end{figure}

For a more quantitative assessment, we show in Fig. \ref{fig:comp_valid} some computed and measured data of $T_{\rm c}$ of some notable superconductors, obtained from the same atomic structure under the same pressure. Four materials with highest-$T_{\rm c}$ shown in this Figure are MgB$_2$ at ambient pressure,\cite{nagamatsu2001superconductivity,margine2013anisotropic,choi2009anisotropic, calandra2010adiabatic, kong2001electron,bohnen2001phonon,choi2002first} H$_3$S at 200 GPa,\cite{Drozdov15,errea2015high} YH$_6$ at 166 GPa,\cite{troyan2021anomalous} and LaYH$_{20}$ at 180 GPa.\cite{semenok2021superconductivity} For each of them, multiple computational schemes were used, involving different values of $\mu^*$ and whether or not the anharminicity is included. These empirical treatments allow the computed $T_{\rm c}$ to spread over a sizable range, i.e., $\simeq 29$K for MgB$_2$, $\simeq 125$K for H$_3$S, $\simeq 100$K for YH$_6$, and $\simeq 70$K for LaYH$_{20}$. Considering the significant challenges in both experimental and computational techniques, the moderate agreement is reasonable and understandable, highlighting important cautions for future works.

Some emerging techniques may be useful for these challenges. Given that the atomic structure predictions can be effectively coupled with powerful ML potentials \cite{gu2019superconducting, yang2021hard, dolui2024feasible, ishikawa2024evolutionary, ferreira2023search, lucrezi2023quantum}, they are expected to be further accelerated by the next developments of this fast-evolving field \cite{Behler:NNPot, Bartok:GAP1, bartok2017machine,Thompson2015316, Chen:MoFF,shapeev2016moment, gubaev2019accelerating,zhang2018deep}. Further into the future, when deep-learning can be used to accelerate the essential quantum mechanics-based calculations like DFT \cite{MLDFT,li2022deep,scherbela2022solving, del2023deep, del2020efficient} and DFPT \cite{li2024deep} in practice, significant advancements may be envisioned. On the other hand, some advanced physics-based computational approaches are also under developments. Among them are SuperConducting DFT \cite{luders2005ab, marques2005ab, sanna2020combining}, a fully {\it ab-initio} approach for $T_{\rm c}$, and a GW perturbation theory based scheme for computing the electron-phonon coupling \cite{li2024electron}. Once these methods can be fully demonstrated, further advances in this field may be anticipated.

\subsection{Materials informatics approaches}
Different from the experimental and theoretical efforts with a century of history and the computational approaches with two decades of blossoming, materials informatics has less that a decade of development in superconductivity understanding and discovery. Compared to the DFPT-based computational approaches, the driving force behind {\it almost all of the discoveries} of hydride superconductors discussed in Sec. \ref{sec:disc}, materials informatics remains in its early stage, both in terms of the methodology development and the impacts to the field. Nevertheless, opportunities exist on the horizon for the informatics-based methods to efficiently complement the traditional (computational and experimental) approaches. Toward this future, some critical challenges, discussed below, are waiting to be resolved.

\subsubsection{Data generation and accumulation}\label{sec:data2}
Data development is the most critical challenge of materials informatics approaches in superconductor discovery. In this area, available data are scarce and insufficient. The biggest superconductor dataset is \texttt{SuperCon}$^2$, developed in 2023 with $\simeq 40,000$ records. The most widely-used dataset, i.e., \texttt{NIMS SuperCon}, in its cleaned version (\texttt{SuperCon}), has $\simeq 16,000$ records \cite{stanev2018machine}. For both of them, only the chemical formula is available for the description of the materials. Attempts to augment the chemical formula with atomic details using structure matching/lookup method produced datasets of up  to $\simeq 5,700$ records in size \cite{zhang2022machine}, as they rely on \texttt{NIMS SuperCon} for the experimental values of $T_{\rm c}$. Such datasets are small and less informative compared to others branches of materials informatics.

Beside a few high-throughput computational efforts that can generate hundreds to thousands of data records of $\lambda$, $\omega_{\rm log}$, and computed $T_{\rm c}$ \cite{shipley2021high, choudhary2022designing, wines2023high, cerqueira2024sampling, hai2023superconductivity}, the vast majority of the literature can only report results from structure predictions works performed for a handful of materials. This is the main reason limiting the available computational data in the field. Initial attempt to collect atomic-level data from literature \cite{tran2023machine} returns nearly 600 records, but more efficient and scalable methods are needed. Given some known limits of the computational approaches, computed data, when developed, should be combined with (more trustworthy) experimental data and ``machine'' learned in a collective and complementing manner.

Scientific literature also hosts a huge, yet essentially untouched, experimental data reservoir. During the last decade, NLP-based approaches such as NER have demonstrated their power in creating \texttt{SC-CoMIcs} \cite{yamaguchi2020sc, yamaguchi2022superconductivity}, \texttt{SuperMat} \cite{foppiano2021supermat}, and \texttt{SuperCon$^2$} \cite{foppiano2023automatic}. Significant future developments are anticipated for these methods can be used to extract the crystallographic information, typically distributed throughout the full-text and in Tables, Supporting Information material, and, in few cases, in some atomic structure formats. In such a challenging endeavor, Large Language Models (LLM) like ChatGPT \cite{chatgpt} and Meta Llama 2 and 3 \cite{llama} could be useful, as recently demonstrated \cite{dagdelen2024structured, zheng2023chatgpt}. Nevertheless, enormous challenges are envisioned for this approach to become mature and efficient in literature data extraction.

\subsubsection{Deep-learning for the physics of superconductivity}\label{sec:deeplearning}
Superconductivity is highly collective and non-local in nature. The coherence length of the Cooper pairs can range from a few nanometers as in LaH$_{10}$ \cite{drozdov2019superconductivity}, YH$_6$, YH$_9$ \cite{kong2021superconductivity} and CeH$_9$ \cite{chen2021high} to several hundred nanometers as in many known superconductors \cite{cadden2009cooper}. Chemical formulae do not include such information, and thus, are insufficient for informatics techniques. Even when the atomic structures are used to describe the materials, recognizing such behaviors is impossible for local-environment- and fragment-based featurzing schemes like SOAP \cite{de2016comparing} and SiRMS \cite{kuz2008hierarchical}. In such the ``hand-crafted'' featurizing schemes, the typical ``neighboring'' distance cutoff of $\sim 10$\AA~ is far too small for the physics of superconductivity. However, bringing this length scale to the order of hundreds of nanometers will certainly (and exponentially) blow up the size and the complexity of feature vectors, making them impractical.

In this context, deep-learning techniques that can directly accept the atomic structures (of course, when data are available) to understand both the local atomic environments and the long-range orders of the materials, could be useful. Deep Neural Network based architectures \cite{konno2021deep} like ATCNN \cite{zeng2019atom} and ALIGNN \cite{choudhary2021atomistic} have been developed and used in some cases with encouraging results. Nevertheless, further developments for this challenging problem, going beyond these initial steps, are needed and anticipated. Equivariant Neural Networks, in which essential physics-inspired invariances are respected, starts emerging and shows its applicability in materials informatics \cite{batzner2022,kaba2022equivariant,gong2023general}.

\subsection{ML-guided search for ambient pressure superconductors}
After two decades being busy with discoveries at hundreds of GPa, recent attentions are gradually shifted to the search for possible high-$T_{\rm c}$ superconductivity at lower pressures \cite{hutcheon2020predicting, cerqueira2024sampling, dolui2024feasible, sanna2024prediction, choudhary2022designing, tran2023machine}. Given the infinite materials space and the complexity of the traditional physics-based approaches, rapid ML models for superconductivity predicting, as discussed in Sec. \ref{sec:ML}, could be useful, but not without challenges. Adding to the challenges outlined in Secs. \ref{sec:data2} and \ref{sec:deeplearning} for developing the ML models, another notable problem is that inferences made outside, or far from, the scope of their experience, are generally not good \cite{balestriero2021learning}. Thus, if a ML model is trained on the superconductors whose $T_{\rm c} \leq 150$ K (at ambient pressure), it may not be able to recognize superconductors with higher $T_{\rm c}$.

One possible way to expand the ``domain of applicability'' is to include in the training data the atomic-level details of the (computational and experimental) discoveries at any pressures, for which $T_{\rm c}$ could be much higher (e.g., $\simeq 200$ K for H$_3$S and $\simeq 250$ K for LaH$_{10}$) \cite{tran2023machine}. Trained on such a dataset, the obtained models may recognize the atomic details related to high-$T_{\rm c}$ superconductivity at any pressures, including ambient pressure. In this approach, external pressure $P$ is assumed to connect with the superconductivity indirectly, i.e., $P$ determines the atomic structures, which, in turn, determines the superconductivity. This assumption has its roots in thermodynamics, where $P$, a macroscopic concept, determines the atomic structures through $P = -\partial H/\partial V$, given that both the enthalpy $H$ and the volume $V$ of an unit cell are solely functions of the atomic structures. The current edition of such a training data, i.e., \texttt{CompSC}, is small, containing 587 atomic structures with 584 values of $\lambda$ and 567 values of $\omega_{\rm log}$ \cite{tran2023machine}. When a substantially bigger and more diverse version is available and advanced deep-learning techniques are developed, the ML models trained on this dataset may be more effective and useful.

Along another dimension, potential non-hydrogen high-$T_{\rm c}$ superconductors have not been explored appropriately. Although hydrides may have favorable vibrational frequency spectrum for high-$T_{\rm c}$ superconductors, this field remains big for other enigmas and wonders. As \texttt{CompSC} currently contains mostly hydrides,\cite{tran2023machine} the inclusion of non-hydrogen superconductors would be critical towards an exploration beyond the hydride-related boundary.

\section{Summary}
Research efforts devoted to superconductor discovery in the last two decades are massive and incredible. Among thousands of superconductors predicted computationally at hundreds of GPa, a few dozens were synthesized and characterized experimentally. The inspiring results have somehow rekindled the dream of room-temperature superconductors. Nevertheless, this ``holy grail'' remains far from being attainable while technical challenges are numerous and enormous. Many of them are related to conducting measurements at hundreds of GPa in a DAC, and then analyzing/interpreting the obtained data \cite{hirsch2021unusual, wang2021absence, hirsch2021nonstandard, gubler2022missing, eremets2022high, hirsch2022superconducting, xing2023observation, chen2023muted}. Theoretical foundations and computational approaches, on the other hand, have their own inherent hard limits that are not easy to overcome.

Materials informatics approaches, emerging as a new frontier of materials research, could be useful to complement the traditional approaches. In the last decade, some essential components of materials informatics for superconductor discovery have been developed, reaching an inspiring level of maturity. Toward the future, tremendous opportunities exist when enormous challenges can be resolved. The most notable challenges are, but not limited to, those related to data generation and curation, deep learning techniques that can capture the physics of superconductivity, and a reliable ML-guided search protocol for high-$T_{\rm c}$ superconductors at ambient pressures. Looking at other branches of materials informatics where tremendous advancements have been made, we believe that superconductor discovery may be significantly advanced with the new developments in this new frontier of materials research.

\section*{Acknowledgements}
Some data used for Figs. \ref{fig:disc} were prepared with support from the San Diego Supercomputer Center (Expanse) within the XSEDE/ACCESS allocation number DMR170031. Work by VNT is supported by the Vietnam National Foundation for Science and Technology Development under Grant 103.01-2021.12.



%
\end{document}